\documentclass[%
preprint
 amsmath,amssymb,
 aps, physrev,
]{revtex4-2}

\usepackage{graphicx}
\usepackage{dcolumn}
\usepackage{bm}
\usepackage{colortbl}
\usepackage{xcolor}
\usepackage{hyperref}
\usepackage{rotating}
\usepackage{float}

\usepackage{ulem}

\begin{document}


\title{\textbf{Large scale statistically validated comorbidity networks}}

\author{Paride Crisafulli}
 \affiliation{IFISC, Instituto de Fisica Interdisciplinar y Sistemas Complejos (CSIC-UIB), Campus Universitat de les Illes Balears, 07122 Palma de Mallorca, Spain}
 
\author{Tobias Galla}
 \affiliation{IFISC, Instituto de Fisica Interdisciplinar y Sistemas Complejos (CSIC-UIB), Campus Universitat de les Illes Balears, 07122 Palma de Mallorca, Spain}

\author{Antti Karlsson}
 \affiliation{Auria Biobank, Turku, Finland.}

\author{Salvatore Miccich\`e}
 \affiliation{Dipartimento di Fisica e Chimica Emilio Segrè, Università degli Studi di Palermo, Palermo, Italy}

\author{Jyrki Piilo}
 \affiliation{Department of Physics and Astronomy, University of Turku, Turku, Finland.}

\author{Rosario N. Mantegna}
 \affiliation{Dipartimento di Fisica e Chimica Emilio Segrè, Università degli Studi di Palermo, Palermo, Italy}
 \affiliation{Complexity Science Hub, Vienna, Austria}

\date{\today}

\begin{abstract}
We obtain comorbidity networks starting from medical information stored in electronic health records collected by the Wellbeing Services County of Southwest Finland (Varha). Based on the data, we associate each patient to one or more diseases and construct complex comorbidity networks associated with large patient cohorts characterized by an age interval and sex. The information about diseases in electronic health records is coded using the highest granularity present in the international classification of diseases (ICD codes) provided by the World Health Organization. We statistically validate links in each cohort’s comorbidity network and furthermore partition the networks into communities of diseases. These are characterized by the over-expression of a few disease categories, and communities from different age or sex cohorts show various similarities in terms of these disease classes. Moreover, all the detected communities for all the cohorts can be organized into a hierarchical tree. This allows us to observe a number of clusters of communities —  originating from diverse age and sex cohorts — that group together communities characterized by the same disease classes. We also perform a dismantling procedure of statistically validated comorbidity networks to highlight those categories of diseases that are most responsible for the compactedness of the comorbidity networks for a given cohort of patients.
\end{abstract}

\keywords{Electronic Health Records, Comorbidity, Complex networks, Statistically Validated Networks}

\maketitle


\section{Introduction} 

Network modeling is a powerful and flexible tool for the investigation of complex systems \cite{newman2018networks,barabasi2016network,latora2017complex}. The investigation of complex networks of biological and/or medical interest has led to research fields referred to as `network medicine' \cite{barabasi2011network} and `network physiology' \cite{ivanov2021new}. One line of research of network medicine concerns comorbidity networks. 

Comorbidity networks are networks of diseases (or conditions -- we will use both terms synonymously). Nodes represent diseases, and the presence of a link between a pair of nodes indicates the co-occurrence of both diseases in the medical history of a patient.  Comorbidity networks can be constructed from information in electronic health records (EHRs) of large health organizations providing health services to patients of regions or countries. 

EHRs originating from the hospitalization process are standardized in most of the countries. Diseases are coded using the international classification of diseases (ICDs) published by the World Health Organization (WHO). This classification is used worldwide to estimate morbidity and mortality statistics. The classification is periodically updated, and it is aimed at facilitating international comparability. The current version is ICD-11. Other versions that are still being used are ICD-9 and ICD-10 \cite{icd-10}.

Since the early work investigating Medicare claims \cite{lee2008implications,hidalgo2009dynamic}, comorbidity networks have been investigated by using small \cite{folino2010comorbidity,roque2011using} and large \cite{chmiel2014spreading,jensen2014temporal,jeong2017network} sets of EHRs of different regional areas and countries. A comparative study of comorbidity patterns in China and the UK has been reported recently \cite{bao2023exploring}. Comorbidity networks are often obtained for specific cohorts of patients; typical stratification is in terms of age and sex. Results for specific cohorts of patients can contribute to devising cohort-specific medical protocols.

In the present study, we analyze large comorbidity networks starting from the dataset of EHRs collected by the Wellbeing Services County of Southwest Finland (Varha) \cite{varha}. These data can be accessed for research purposes via Auria Clinical Informatics \cite{auria}. Auria Clinical Informatics is a company situated in Turku, Finland, interacting with Turku University Hospital and the University of Turku and operating under license from Valvira, the National Supervisory Authority for Welfare and Health in Finland.  

Previous studies of comorbidity networks 
\cite{hidalgo2009dynamic,folino2010comorbidity,roque2011using,chmiel2014spreading,jensen2014temporal,jeong2017network,fotouhi2018statistical,bao2023exploring}
have been performed by considering ICD classification at so-called ``level 3" (i.e., a classification level with a set of 3 characters as, for example I10). Here we investigate different cohorts of EHR of Finnish patients by using a ICD classification at ``level 4" (i.e., with a resolution of four characters as, for example F32.2). Level-3 codes are used for high-level categorization in clinical diagnosis or health statistics while a level-4 classification provides a more specific diagnosis, often used by healthcare providers and researchers to determine exact treatments and prognosis. Level-4 classification is therefore more detailed and closer to the practice of medical doctors. By moving from  level 3 to level 4 we improve the resolution of the comorbidity network description. This choice allows us to analyze each disease at the highest level of description that is present in medical records. Further details can be found in Sec.~\ref{sec:icd}.

In our analysis, we start from a bipartite patient-disease network in which all diagnosis are recorded as links between a patient and the conditions the patient has been diagnosed with. We then obtain a projected network comprising of only disease nodes, each representing one level-4 ICD code. Two ICD codes are connected in this network if there is at least patient in the target cohort who has been diagnosed with both conditions. We refer to this network as `PROJ' (for projected network). 

We assume that not all links in this projected network are medically informative to the same degree. 
To highlight comorbidity relations that are not explainable as arising from the prevalence of diseases, for each link in the PROJ network, we perform a statistical validation \cite{svn-paper,li2014statistically,hatzopoulos2015quantifying, serrano2009extracting,radicchi2011information,saracco2015randomizing,saracco2017inferring,marcaccioli2019polya,kobayashi2019structured,vallarano2021fast}. For each link in the original PROJ network, we assess if the link is compatible with a null hypothesis of random co-occurrence, taking into account the prevalences of the different conditions in the patient cohort for which the network is being constructed. 
 If a link is compatible with random co-occurrence to a specified level of statistical confidence, we remove the link. By performing this statistical test for all links in the original PROJ network, we extract what we will refer to as the `statistically validated network' (SVN) of diseases. This network contains a subset of all co-occurrences of ICD codes in the dataset.
For example, E03.9 (Hypothyroidism, unspecified) and J01.0 (Acute maxillary sinusitis) are common diseases in the 50-59 cohort of males, so it’s quite likely that several patients got both diseases just due to their high prevalence. In fact, in the 50-59 cohort, these two diseases are connected in the PROJ network, but they are not connected in the corresponding SVN suggesting that the medical relevance of this observed comorbidity in some patients might be limited or even absent. The validation procedure implies the execution of a large number of tests (one for each link in the original PROJ network), requiring a multiple-hypothesis test correction~\cite{rupert2012simultaneous}. Specifically, we used the so-called `control of the false discovery rate' \cite{fdr-corr}. 
 
Having constructed the PROJ networks and the SNVs for different cohorts of patients, we search communities of diseases in both types of network using community-detection methods for complex networks \cite{fortunato2016community}.
This provides communities of diseases characterized by pronounced intra-community connectivity. While the original PROJ networks are so dense that Infomap, a widely used community detection algorithm \cite{rosvall2008maps}, is unsuccessful in detecting distinct clusters of diseases, the same algorithm finds several clusters of diseases in the SVNs. The communities have sizes ranging from small to medium to large number of ICD codes, and we find that they are informative about different groups of comorbidities for different cohorts of patients.

In fact, we verify that different cohorts of patients are characterized by distinct sets of communities in the SVNs. We provide a comparison of comorbidity communities in different patient cohorts by computing a similarity measure between them and grouping them by using agglomerative hierarchical clustering \cite{kaufman2009finding}. The inspection of the hierarchical tree is supported by a statistical test detecting the category or categories of diseases with an over-expressed number in each community (see Section \ref{sec:community-over-exp}). With this characterization, we note that communities in which the same disease classes are over-expressed tend to group in the same subregions of the hierarchical tree, even when they belong to cohorts of patients of different age and sex.

We also provide a specific example of ICD community analysis by discussing in some detail communities with an over-expressed presence of mental and behavioral disorders. A similar detailed analysis can be performed for any disease category of interest.

Healthcare policy decisions focused on specific cohorts of the patient population benefit from information about comorbidity relationships specific to those cohorts. In different cohorts, different diseases might be characterized by a high value of node betweenness, making these diseases crucial to sustain the structure of the main component of the comorbidity network. To highlight the categories of these diseases, we perform an investigation of ICD nodes contributing to the robustness of comorbidity PROJ networks and SVNs. This analysis would allow us to highlight what are the ICD code categories that have the most prominent role in keeping the comorbidity network cohesive.

Specifically, we perform a so-called `dismantling procedure' \cite{dism,wandelt2018comparative} for the PROJ networks and the SVNs. 
Network dismantling mimics preventive medicine approaches applied to different cohorts of patients aimed at the minimization of medical comorbidity. 
We model network dismantling by a successive removal of ICD nodes from comorbidity networks, aiming at fragmenting the network into smaller, disconnected components.  Using this procedure, we highlight some prominent roles for specific categories of disease in determining the compactness of the comorbidity network. We verify that this information is cohort-specific several cohorts characterized by age and sex.

The remainder of the paper is organized as follows. In Section \ref{sect:data} we present the data and the methodology used to obtain comorbidity networks together with some summary information about the PROJ networks and SVNs. Section \ref{sec:svn-props} presents briefly the basic properties of the SVNs while in Section \ref{sec:communities} we discuss the disease communities detected in the SVNs. Section \ref{sect:case_study} contains two case studies on mental and behavioral disorders, and in Section \ref{sect:dismantling} we describe results obtained by applying a dismantling procedure to PROJ networks and SVNs of different cohorts of patients. In Section \ref{sect:conclusions} we present our conclusions.

\section{Data and methodology}  \label{sect:data} 

\subsection{Availability of data materials}

The data investigated in this study are proprietary data of Auria Clinical Informatics which operates in connection with Varha. Data can be accessed with permission from Varha. The present study analyzes disease networks obtained with the approval of the Institutional Review Board of Turku University Hospital (license number T152/2017 \cite{paripaper}). Informed consent was waived due to the study's retrospective design, according to Finnish legislation on the secondary use of health data.

\subsection{Preprocessing and cohorting of medical data}\label{sec:icd}

To classify medical conditions we use the International Classification of Diseases, 10th Revision (ICD-10). This is a globally recognized classification system developed by the WHO for classifying and coding diseases, health conditions, and causes of death. It provides standardized codes that facilitate the documentation, reporting, and analysis of health data across different regions and healthcare settings. Starting from the 10th revision, ICD codes consist of a letter (disease class) followed by two or more digits (which can be separated by decimal points), progressively deepening the identification of the disease (or condition for a subset of categories). For example, codes starting with the letter F refer to mental and behavioral disorders. Within this group, F32 (i.e., a so-called level 3 code composed by three characters) represents the occurrence of depressive episodes, and further, F32.2 (i.e., a level 4 code) stands for severe depressive episode without psychotic symptoms. The full list of Level 4 codes can be found on the WHO website \cite{icd-10}. In Table \ref{tab:letters} we list the disease categories classified by the first letter of ICD codes. In our study,  we set the granularity of our study by using leve 4 ICD codes. This defines a set of 9303 different codes.

Data covers the period between the 1st of January 2004 and the 31st of July 2019, i.e., for a time period covering 16 years, and includes a total of 628,831 patients. Each of the more than 20 million line records in the dataset consists of an anonymized patient ID, the ICD code of the diagnosis, the timestamp of the visit, and the patient's metadata, such as age at the time of the visit and sex. Data entries include each diagnosis registered in the aforementioned time window in a hospital connected to Varha, regardless of the fact that the patient had been treated or not. 

In a first step, we identify all entries for a specific patient and thus construct a list of all conditions this patient is diagnosed with (at any time). For further analysis, we divide patients into cohorts characterized by sex and age. Specifically, we study 10-year age groups, resulting in a total of 18 different cohorts (9 age groups, two sexes). The reference age is always the age at the time of diagnosis. For a given sex and age group, we include all conditions patients were diagnosed with within the age limits of the cohort or earlier. This is because many diseases can have long-term effects, causing comorbidities. For example, if a patient was diagnosed for the first time with ICD code S61.9 (open wound of wrist and hand, part unspecified) when he or she was 17 years old and has a first-time diagnosis with ICD code F50.1 (atypical anorexia nervosa) at age 24, then ICD code S61.9 contributes to the 10-19 cohort, and both diagnoses (S61.9 and F50.1) contribute to the 20-29 age cohort.

\begin{table*}
\begin{tabular}{|c|l|}
\hline
\textbf{A-B}   & Certain infectious and parasitic diseases                                               \\ \hline
\textbf{C}     & Neoplasms                                                                               \\ \hline
\textbf{D}     & Neoplasms, diseases of the blood and blood-forming organs and certain immune disorders  \\ \hline
\textbf{E}     & Endocrine, nutritional and metabolic diseases                                           \\ \hline
\textbf{F}     & Mental and behavioural disorders                                                        \\ \hline
\textbf{G}     & Diseases of the nervous system                                                          \\ \hline
\textbf{H}     & Diseases of the eye and adnexa,  diseases of the ear and mastoid process                \\ \hline
\textbf{I}     & Diseases of the circulatory system                                                      \\ \hline
\textbf{J}     & Diseases of the respiratory system                                                      \\ \hline
\textbf{K}     & Diseases of the digestive system                                                        \\ \hline
\textbf{L}     & Diseases of the skin and subcutaneous tissue                                            \\ \hline
\textbf{M}     & Diseases of the musculoskeletal system and connective tissue                            \\ \hline
\textbf{N}     & Diseases of the genitourinary system                                                    \\ \hline
\textbf{O}     & Pregnancy, childbirth and the puerperium                                                \\ \hline
\textbf{P}     & Certain conditions originating in the perinatal period                                  \\ \hline
\textbf{Q}     & Congenital malformations, deformations and chromosomal abnormalities                    \\ \hline
\textbf{R}     & Symptoms, signs and abnormal clinical and laboratory findings, not elsewhere classified \\ \hline
\textbf{S-T}   & Injury, poisoning and certain other consequences of external causes                     \\ \hline
\textbf{U}     & Codes for special purposes                                                              \\ \hline
\textbf{V-X-Y} & External causes of morbidity and mortality                                              \\ \hline
\textbf{Z}     & Factors influencing health status and contact with health services                      \\ \hline
\end{tabular}
\caption{Definition of each ICD category (labeled by a single letter). Definitions are taken from the ICD-10 website \cite{icd-10}}
\label{tab:letters}
\end{table*}

\subsection{Bipartite and projected networks}

\begin{table*}
\scalebox{0.88}{
\begin{tabular}{llllllllll}
\hline
\rowcolor[HTML]{CDE2FF} 
\multicolumn{1}{|l|}{\cellcolor[HTML]{CDE2FF}\textbf{Number of}}    & \multicolumn{1}{l|}{\cellcolor[HTML]{FFCCC9}\textbf{0-9 F}}   & \multicolumn{1}{l|}{\cellcolor[HTML]{CDE2FF}\textbf{0-9 M}}   & \multicolumn{1}{l|}{\cellcolor[HTML]{FFCCC9}\textbf{10-19 F}} & \multicolumn{1}{l|}{\cellcolor[HTML]{CDE2FF}\textbf{10-19 M}} & \multicolumn{1}{l|}{\cellcolor[HTML]{FFCCC9}\textbf{20-29 F}} & \multicolumn{1}{l|}{\cellcolor[HTML]{CDE2FF}\textbf{20-29 M}} & \multicolumn{1}{l|}{\cellcolor[HTML]{FFCCC9}\textbf{30-39 F}} & \multicolumn{1}{l|}{\cellcolor[HTML]{CDE2FF}\textbf{30-39 M}} & \multicolumn{1}{l|}{\cellcolor[HTML]{FFCCC9}\textbf{40-49 F}} \\ \hline
\rowcolor[HTML]{CDE2FF} 
\multicolumn{1}{|l|}{\cellcolor[HTML]{CDE2FF}Patients}    & \multicolumn{1}{l|}{\cellcolor[HTML]{FFCCC9}46387}            & \multicolumn{1}{l|}{\cellcolor[HTML]{CDE2FF}53130}            & \multicolumn{1}{l|}{\cellcolor[HTML]{FFCCC9}44451}            & \multicolumn{1}{l|}{\cellcolor[HTML]{CDE2FF}45549}            & \multicolumn{1}{l|}{\cellcolor[HTML]{FFCCC9}71057}            & \multicolumn{1}{l|}{\cellcolor[HTML]{CDE2FF}55076}            & \multicolumn{1}{l|}{\cellcolor[HTML]{FFCCC9}64337}            & \multicolumn{1}{l|}{\cellcolor[HTML]{CDE2FF}50887}            & \multicolumn{1}{l|}{\cellcolor[HTML]{FFCCC9}57487}            \\ \hline
\rowcolor[HTML]{CDE2FF} 
\multicolumn{1}{|l|}{\cellcolor[HTML]{CDE2FF}Bipartite links} & \multicolumn{1}{l|}{\cellcolor[HTML]{FFCCC9}192414}           & \multicolumn{1}{l|}{\cellcolor[HTML]{CDE2FF}247572}           & \multicolumn{1}{l|}{\cellcolor[HTML]{FFCCC9}236009}           & \multicolumn{1}{l|}{\cellcolor[HTML]{CDE2FF}237602}           & \multicolumn{1}{l|}{\cellcolor[HTML]{FFCCC9}470116}           & \multicolumn{1}{l|}{\cellcolor[HTML]{CDE2FF}236183}           & \multicolumn{1}{l|}{\cellcolor[HTML]{FFCCC9}583306}           & \multicolumn{1}{l|}{\cellcolor[HTML]{CDE2FF}240946}           & \multicolumn{1}{l|}{\cellcolor[HTML]{FFCCC9}469684}           \\ \hline
\rowcolor[HTML]{CDE2FF} 
\multicolumn{1}{|l|}{\cellcolor[HTML]{CDE2FF}PROJ nodes} & \multicolumn{1}{l|}{\cellcolor[HTML]{FFCCC9}3904}             & \multicolumn{1}{l|}{\cellcolor[HTML]{CDE2FF}4121}             & \multicolumn{1}{l|}{\cellcolor[HTML]{FFCCC9}5250}             & \multicolumn{1}{l|}{\cellcolor[HTML]{CDE2FF}4933}             & \multicolumn{1}{l|}{\cellcolor[HTML]{FFCCC9}5904}             & \multicolumn{1}{l|}{\cellcolor[HTML]{CDE2FF}5163}             & \multicolumn{1}{l|}{\cellcolor[HTML]{FFCCC9}6113}             & \multicolumn{1}{l|}{\cellcolor[HTML]{CDE2FF}5296}             & \multicolumn{1}{l|}{\cellcolor[HTML]{FFCCC9}6184}             \\ \hline
\rowcolor[HTML]{CDE2FF} 
\multicolumn{1}{|l|}{\cellcolor[HTML]{CDE2FF}PROJ links} & \multicolumn{1}{l|}{\cellcolor[HTML]{FFCCC9}215096}           & \multicolumn{1}{l|}{\cellcolor[HTML]{CDE2FF}266002}           & \multicolumn{1}{l|}{\cellcolor[HTML]{FFCCC9}407770}           & \multicolumn{1}{l|}{\cellcolor[HTML]{CDE2FF}381156}           & \multicolumn{1}{l|}{\cellcolor[HTML]{FFCCC9}711053}           & \multicolumn{1}{l|}{\cellcolor[HTML]{CDE2FF}398215}           & \multicolumn{1}{l|}{\cellcolor[HTML]{FFCCC9}911158}           & \multicolumn{1}{l|}{\cellcolor[HTML]{CDE2FF}467127}           & \multicolumn{1}{l|}{\cellcolor[HTML]{FFCCC9}964727}           \\ \hline
\rowcolor[HTML]{CDE2FF} 
\multicolumn{1}{|l|}{\cellcolor[HTML]{CDE2FF}SVN nodes}   & \multicolumn{1}{l|}{\cellcolor[HTML]{FFCCC9}1531}             & \multicolumn{1}{l|}{\cellcolor[HTML]{CDE2FF}1810}             & \multicolumn{1}{l|}{\cellcolor[HTML]{FFCCC9}2401}             & \multicolumn{1}{l|}{\cellcolor[HTML]{CDE2FF}2276}             & \multicolumn{1}{l|}{\cellcolor[HTML]{FFCCC9}3374}             & \multicolumn{1}{l|}{\cellcolor[HTML]{CDE2FF}2437}             & \multicolumn{1}{l|}{\cellcolor[HTML]{FFCCC9}3503}             & \multicolumn{1}{l|}{\cellcolor[HTML]{CDE2FF}2549}             & \multicolumn{1}{l|}{\cellcolor[HTML]{FFCCC9}3632}             \\ \hline
\rowcolor[HTML]{CDE2FF} 
\multicolumn{1}{|l|}{\cellcolor[HTML]{CDE2FF}SVN links}   & \multicolumn{1}{l|}{\cellcolor[HTML]{FFCCC9}9192}             & \multicolumn{1}{l|}{\cellcolor[HTML]{CDE2FF}11629}            & \multicolumn{1}{l|}{\cellcolor[HTML]{FFCCC9}14861}            & \multicolumn{1}{l|}{\cellcolor[HTML]{CDE2FF}13652}            & \multicolumn{1}{l|}{\cellcolor[HTML]{FFCCC9}31306}            & \multicolumn{1}{l|}{\cellcolor[HTML]{CDE2FF}13178}            & \multicolumn{1}{l|}{\cellcolor[HTML]{FFCCC9}37204}            & \multicolumn{1}{l|}{\cellcolor[HTML]{CDE2FF}16624}            & \multicolumn{1}{l|}{\cellcolor[HTML]{FFCCC9}37187}            \\ \hline
\textbf{}                                                                &                                                               &                                                               &                                                               &                                                               &                                                               &                                                               &                                                               &                                                               &                                                               \\ \hline
\rowcolor[HTML]{FFCCC9} 
\multicolumn{1}{|l|}{\cellcolor[HTML]{FFCCC9}\textbf{Number of}}          & \multicolumn{1}{l|}{\cellcolor[HTML]{CDE2FF}\textbf{40-49 M}} & \multicolumn{1}{l|}{\cellcolor[HTML]{FFCCC9}\textbf{50-59 F}} & \multicolumn{1}{l|}{\cellcolor[HTML]{CDE2FF}\textbf{50-59 M}} & \multicolumn{1}{l|}{\cellcolor[HTML]{FFCCC9}\textbf{60-69 F}} & \multicolumn{1}{l|}{\cellcolor[HTML]{CDE2FF}\textbf{60-69 M}} & \multicolumn{1}{l|}{\cellcolor[HTML]{FFCCC9}\textbf{70-79 F}} & \multicolumn{1}{l|}{\cellcolor[HTML]{CDE2FF}\textbf{70-79 M}} & \multicolumn{1}{l|}{\cellcolor[HTML]{FFCCC9}\textbf{80+ F}}   & \multicolumn{1}{l|}{\cellcolor[HTML]{CDE2FF}\textbf{80+ M}}   \\ \hline
\rowcolor[HTML]{FFCCC9} 
\multicolumn{1}{|l|}{\cellcolor[HTML]{FFCCC9}Patients}    & \multicolumn{1}{l|}{\cellcolor[HTML]{CDE2FF}52821}            & \multicolumn{1}{l|}{\cellcolor[HTML]{FFCCC9}66699}            & \multicolumn{1}{l|}{\cellcolor[HTML]{CDE2FF}62776}            & \multicolumn{1}{l|}{\cellcolor[HTML]{FFCCC9}65587}            & \multicolumn{1}{l|}{\cellcolor[HTML]{CDE2FF}63156}            & \multicolumn{1}{l|}{\cellcolor[HTML]{FFCCC9}55947}            & \multicolumn{1}{l|}{\cellcolor[HTML]{CDE2FF}48046}            & \multicolumn{1}{l|}{\cellcolor[HTML]{FFCCC9}41505}            & \multicolumn{1}{l|}{\cellcolor[HTML]{CDE2FF}24463}            \\ \hline
\rowcolor[HTML]{FFCCC9} 
\multicolumn{1}{|l|}{\cellcolor[HTML]{FFCCC9}Bipartite links}  & \multicolumn{1}{l|}{\cellcolor[HTML]{CDE2FF}292536}           & \multicolumn{1}{l|}{\cellcolor[HTML]{FFCCC9}506556}           & \multicolumn{1}{l|}{\cellcolor[HTML]{CDE2FF}418254}           & \multicolumn{1}{l|}{\cellcolor[HTML]{FFCCC9}568945}           & \multicolumn{1}{l|}{\cellcolor[HTML]{CDE2FF}541195}           & \multicolumn{1}{l|}{\cellcolor[HTML]{FFCCC9}578651}           & \multicolumn{1}{l|}{\cellcolor[HTML]{CDE2FF}530509}           & \multicolumn{1}{l|}{\cellcolor[HTML]{FFCCC9}539798}           & \multicolumn{1}{l|}{\cellcolor[HTML]{CDE2FF}345711}           \\ \hline
\rowcolor[HTML]{FFCCC9} 
\multicolumn{1}{|l|}{\cellcolor[HTML]{FFCCC9}PROJ nodes} & \multicolumn{1}{l|}{\cellcolor[HTML]{CDE2FF}5561}             & \multicolumn{1}{l|}{\cellcolor[HTML]{FFCCC9}6349}             & \multicolumn{1}{l|}{\cellcolor[HTML]{CDE2FF}5910}             & \multicolumn{1}{l|}{\cellcolor[HTML]{FFCCC9}6130}             & \multicolumn{1}{l|}{\cellcolor[HTML]{CDE2FF}6047}             & \multicolumn{1}{l|}{\cellcolor[HTML]{FFCCC9}5860}             & \multicolumn{1}{l|}{\cellcolor[HTML]{CDE2FF}5713}             & \multicolumn{1}{l|}{\cellcolor[HTML]{FFCCC9}5429}             & \multicolumn{1}{l|}{\cellcolor[HTML]{CDE2FF}4878}             \\ \hline
\rowcolor[HTML]{FFCCC9} 
\multicolumn{1}{|l|}{\cellcolor[HTML]{FFCCC9}PROJ links} & \multicolumn{1}{l|}{\cellcolor[HTML]{CDE2FF}613123}           & \multicolumn{1}{l|}{\cellcolor[HTML]{FFCCC9}1014863}          & \multicolumn{1}{l|}{\cellcolor[HTML]{CDE2FF}832433}           & \multicolumn{1}{l|}{\cellcolor[HTML]{FFCCC9}1083676}          & \multicolumn{1}{l|}{\cellcolor[HTML]{CDE2FF}1020775}          & \multicolumn{1}{l|}{\cellcolor[HTML]{FFCCC9}1051192}          & \multicolumn{1}{l|}{\cellcolor[HTML]{CDE2FF}973948}           & \multicolumn{1}{l|}{\cellcolor[HTML]{FFCCC9}859456}           & \multicolumn{1}{l|}{\cellcolor[HTML]{CDE2FF}682662}           \\ \hline
\rowcolor[HTML]{FFCCC9} 
\multicolumn{1}{|l|}{\cellcolor[HTML]{FFCCC9}SVN nodes}   & \multicolumn{1}{l|}{\cellcolor[HTML]{CDE2FF}2897}             & \multicolumn{1}{l|}{\cellcolor[HTML]{FFCCC9}3805}             & \multicolumn{1}{l|}{\cellcolor[HTML]{CDE2FF}3438}             & \multicolumn{1}{l|}{\cellcolor[HTML]{FFCCC9}3709}             & \multicolumn{1}{l|}{\cellcolor[HTML]{CDE2FF}3632}             & \multicolumn{1}{l|}{\cellcolor[HTML]{FFCCC9}3405}             & \multicolumn{1}{l|}{\cellcolor[HTML]{CDE2FF}3103}             & \multicolumn{1}{l|}{\cellcolor[HTML]{FFCCC9}2574}             & \multicolumn{1}{l|}{\cellcolor[HTML]{CDE2FF}2122}             \\ \hline
\rowcolor[HTML]{FFCCC9} 
\multicolumn{1}{|l|}{\cellcolor[HTML]{FFCCC9}SVN links}   & \multicolumn{1}{l|}{\cellcolor[HTML]{CDE2FF}22068}            & \multicolumn{1}{l|}{\cellcolor[HTML]{FFCCC9}38539}            & \multicolumn{1}{l|}{\cellcolor[HTML]{CDE2FF}32072}            & \multicolumn{1}{l|}{\cellcolor[HTML]{FFCCC9}39181}            & \multicolumn{1}{l|}{\cellcolor[HTML]{CDE2FF}37627}            & \multicolumn{1}{l|}{\cellcolor[HTML]{FFCCC9}34911}            & \multicolumn{1}{l|}{\cellcolor[HTML]{CDE2FF}29397}            & \multicolumn{1}{l|}{\cellcolor[HTML]{FFCCC9}20500}            & \multicolumn{1}{l|}{\cellcolor[HTML]{CDE2FF}12379}            \\ \hline
\end{tabular}
}
\caption{Number of nodes and links in the bipartite, projected (PROJ) , and validated (SVN) networks for different cohorts of patients.}
\label{tab:net-stats}
\end{table*}

Bipartite networks for the different cohorts are built from the dataset by connecting each patient ID with ICD codes reported in her/his medical history if this patient was diagnosed with this condition at least once at an age in the cohort age limits, or earlier. Overall, there is a slight sex imbalance in the number of patients in the datset, 48.1\% are male, and 51.9\% female. Some summary statistics for the PROJ networks and SVNs are shown in Table \ref{tab:net-stats} for the different cohorts. The number of links in the bipartite network is larger in the age cohorts from 50 to 79 years compared to other age groups for both sexes. In the age range 20-39, the number of links in the bipartite network is larger for females than for males due to pregnancy-related diagnoses.

\begin{figure*}
    \centering
        \includegraphics[width=\linewidth]{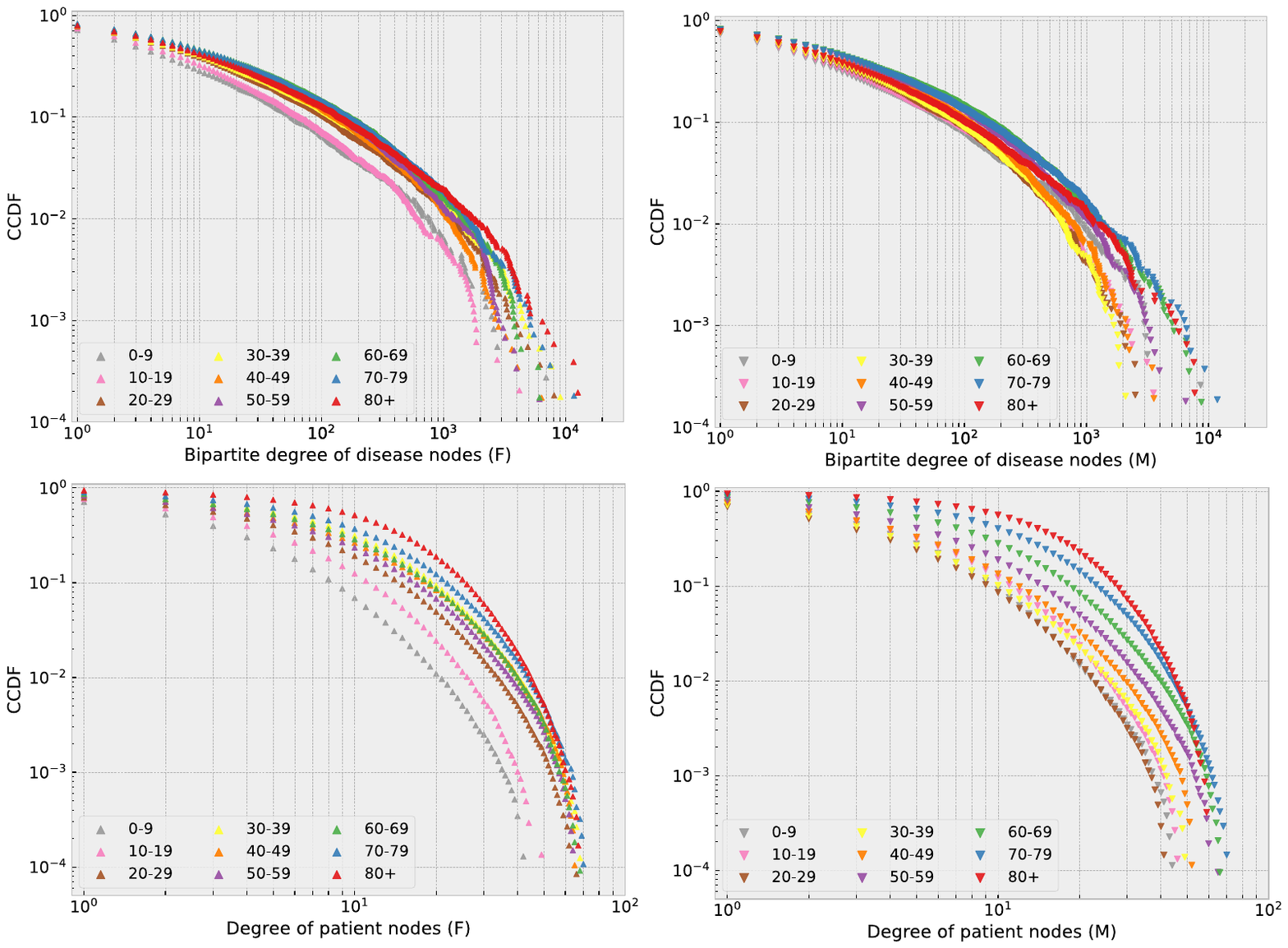}
    \caption{Complementary cumulative distribution function of the degree of ICD nodes (top panels) and of patient nodes (bottom panels) in the bipartite network for all patient cohorts of females (left) and males (right).}
    \label{fig:patient-k}
\end{figure*}

As shown in Fig.~\ref{fig:patient-k}, for all cohorts, the degree of ICD nodes in the bipartite network is highly heterogeneous while the degree of patient nodes is only moderately heterogeneous. The broader nature of the ICD distribution shows the need of a statistical validation.

\subsection{Validating the networks} \label{sec:validation}

We now describe the statistical validation procedure by which we obtain the SVNs from the PROJ networks.

 Suppose the initial bipartite network for a cohort has  $N_D$ disease nodes (ICD codes) and $N_P$ patient nodes, with a link connecting a patient if this patient was diagnosed with the condition at an age in the cohort age period or earlier. The PROJ network then consists of $N_D$ nodes, and a link exists between any two nodes if at least one patient in the cohort had both diseases. 
 
 We focus now on two nodes in the PROJ network, A and B. These are two ICD codes. Assume now that $N_A$ is the number of patients in the cohort diagnosed with condition A, and $N_B$ the number of patients with condition B. Further assume that $N_{AB}$ patients in the cohort have been diagnosed with both conditions (at an age within the age limits of the cohort or earlier). 
 
We now formulate the null hypothesis of random co-occurrence of condition A and B. This can be thought of as follows: there are $N_A$ out of $N_P$ patients with condition A and $N_B$ patients with condition B. We can then ask what the probability is to have a given number $\ell$ of patients with both conditions, if the $N_A$ and $N_B$ individuals were selected at random from the total of $N_P$ patients. Under the assumption that the heterogeneity of number of diseases for patient is moderate the probability of the null hypothesis of observing both conditions is given with a good approximation by \cite{svn-paper,li2014statistically}:
\begin{equation}
H_{AB}(\ell|N_P,N_A,N_B) = \frac{{N_A \choose \ell}{N_P - N_A \choose N_B - \ell}}{{N_P \choose N_B}}.
\label{eq:hg-ab}
\end{equation}
Eq. \ref{eq:hg-ab} approximates to the exact probability to have a given number $\ell$ of patients with both conditions in the absence  of heterogeneity of the degree of the patient's nodes \cite{svn-paper}. Under the hypothesis of random co-occurrence the probability of observing $N_{AB}$ or more co-occurrences is:
\begin{equation}
    p(\ell \geq N_{AB}) = \sum_{\ell = {N_{AB}}}^{{\rm min}(N_A,N_B)} H_{AB}(\ell|N_P,N_A,N_B).
\end{equation}
This allows us to test whether or not the observation of $N_{AB}$ patients in the data with both conditions is compatible with the null hypothesis of random co-occurrence at a given level of statistical confidence. To perform the hypothesis test, we set the univariate statistical threshold at 0.01 and then apply false discovery rate (FDR) correction \cite{fdr-corr} required when performing multiple hypothesis tests. If a link in the PROJ network leads to a $p$-value lower than expected for the FDR correction, we conclude that the null hypothesis is rejected, i.e., the empirical observation of $N_{AB}$ patients with both conditions A and B is not statistically compatible with random co-occurrence hypothesis. The link is then included in the SVN. Conversely, when the null hypothesis is not rejected, we do not include the comorbidity link in the SVN. Finally, isolated nodes are not included in the SVN. The validation was performed by using the Python module accessible at the \texttt{svalnet} github repository \cite{svalnet}.


\section{Results: comorbidity networks}  \label{sect:results}  

\subsection{Properties of the validated networks} \label{sec:svn-props}

Table \ref{tab:net-stats} shows that the number of validated links is less than 5\% of that in the PROJ network. 
This large reduction of comorbidity links does not significantly affect the number of ICD nodes present in the SVNs. In fact, the percent of retained nodes is ranging from about 40\% to 60\% therefore providing information on a large number of ICD lewvel 4 nodes. 
For all cohorts, both for PROJ networks and SVNs consist of a largest connected component (LCC) containing more than 90\% of the network nodes. The remaining nodes are grouped into several small components each with only a small number of nodes.

\begin{figure*} 
    \centering
            \includegraphics[width=\linewidth]{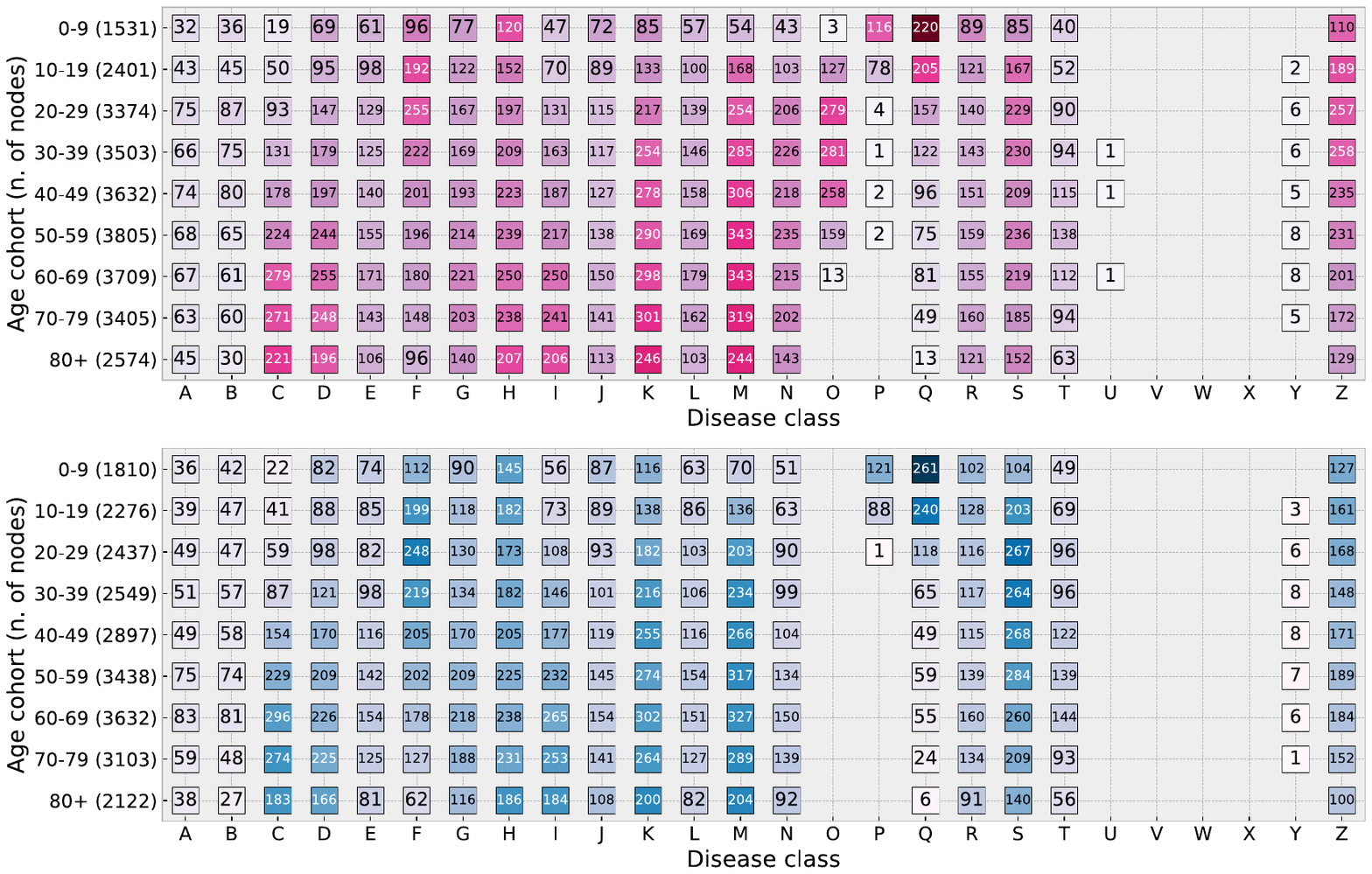}
    \caption{Statistics of ICD category composition of SVNs for all cohorts of patients. ICD categories (columns) for different age cohorts (rows) and sexes (females in the upper panel and males in the lower). The number in brackets next to the age limits of the cohort is the total number of nodes in the SVN. The number inside the squares is the number of nodes of the disease class (the letter in the ICD code) in the given network, while the color is proportional to the fraction of diseases that belong to that class}
    
    \label{fig:letter-frac}
\end{figure*}

For a summary of the statistics of ICD codes present in the SVNs for all cohorts, see Fig. \ref{fig:letter-frac}. In general, the two sexes do not present strong differences in disease distributions, except for specific categories. Prominent examples of differences observed between sexes are trivially the categories O (pregnancy, childbirth and the puerperiuas), and further N (diseases of the genitourinary system). A higher number of nodes of this category are seen for each age-specific female network (Fig. \ref{fig:letter-frac}) compared to the male network for the same age bracket. Categories H, K, and M are present in similar quantities through all ages, while other categories turn out to be more expressed at specific age intervals, e.g. categories P and Q in early ages, F and S in teenagers and young adults, and C, D, and I at later ages. 

\subsection{Disease categories and community structure} \label{sec:communities}

\subsubsection{Community detection in comorbidity networks}
We now analyze the structure of comorbidity networks by looking at subgroups of ICD codes highly connected within each other. To this end, we perform community detection. This is an unsupervised data mining procedure to identify groups in a complex network whose nodes are more connected between each other rather than to other nodes in the network~\cite{fortunato2016community}. We chose the Infomap algorithm \cite{rosvall2008maps} to perform community detection both in the PROJ networks and SVNs. In the PROJ networks, only one very large community is found by the algorithm for each age and sex cohort. In other words, the PROJ network of diseases has a high link density in all regions of the network, and the algorithm is not able to detect distinct regions of the network. On the other hand, in the SVNs many communities are observed for each cohort of patients. 

\subsubsection{Number of disease communities in different cohorts}
%

In Table \ref{tab:comm_svn}, we report the number of ICD communities detected by the Infomap algorithm for all cohorts. Communities vary in size ranging from groups of two ICD codes to communities with hundreds of nodes.  
The fact that the number of communities is bigger than one in each SVN shows how our method highlights more informative subnetworks than the PROJ ones, where Infomap fails to infer distinct communities and returns a single one for each connected component of each cohort. 
The difference in the number of communities between male and female SVNs does not reflect the difference in the number of nodes. This hints to the fact that communities are more influenced by disease class rather than the size of the network. Although medical information can be obtained by the inspection of communities of any size, in the present study, we focus our attention to communities larger than 25 nodes.

\begin{table*}
\begin{tabular}{|c|c|c|c|c|c|}
\hline
\textbf{Cohort} & \textbf{\begin{tabular}[c]{@{}c@{}}Nodes\\ of SVN\end{tabular}} & \textbf{\begin{tabular}[c]{@{}c@{}}\# of comm. \\ (any size)\end{tabular}} & 
\textbf{\begin{tabular}[c]{@{}c@{}}\# of comm.\\ (nodes \textgreater 10)\end{tabular}} & \textbf{\begin{tabular}[c]{@{}c@{}}\# of comm.\\ (nodes \textgreater 25)\end{tabular}} & \textbf{\begin{tabular}[c]{@{}c@{}}Size of the largest\\ SVN community\end{tabular}} \\ \hline
0-9	F	&	1531	&	145	&	24	&	10	&	301		\\	\hline
0-9	M	&	1810	&	152	&	27	&	10	&	384		\\	\hline
Oct-19	F	&	2401	&	178	&	42	&	20	&	233		\\	\hline
Oct-19	M	&	2276	&	202	&	41	&	15	&	331		\\	\hline
20-29	F	&	3374	&	282	&	44	&	23	&	666		\\	\hline
20-29	M	&	2437	&	176	&	46	&	20	&	294		\\	\hline
30-39	F	&	3503	&	284	&	50	&	22	&	1182		\\	\hline
30-39	M	&	2549	&	160	&	56	&	22	&	434		\\	\hline
40-49	F	&	3632	&	253	&	52	&	28	&	373		\\	\hline
40-49	M	&	2897	&	157	&	50	&	26	&	285		\\	\hline
50-59	F	&	3805	&	267	&	46	&	26	&	748		\\	\hline
50-59	M	&	3438	&	227	&	50	&	24	&	482		\\	\hline
60-69	F	&	3709	&	233	&	49	&	27	&	621		\\	\hline
60-69	M	&	3632	&	212	&	48	&	26	&	710		\\	\hline
70-79	F	&	3405	&	237	&	49	&	25	&	984		\\	\hline
70-79	M	&	3103	&	178	&	52	&	20	&	949		\\	\hline
80+	F	&	2574	&	166	&	48	&	20	&	798		\\	\hline
80+	M	&	2122	&	187	&	39	&	16	&	509		\\	\hline
\end{tabular}
\caption{Number of communities of any size, size larger than 10 nodes or 25 nodes of SVNs for each cohort. We also report the number of nodes of the SVN and the number of nodes of the largest community detected. The total number of detected communities of size larger than 25 ICD codes is 380.}
\label{tab:comm_svn}
\end{table*}

\subsubsection{Similarity of communities across cohorts, and hierarchical tree}

One might expect that communities of ICDs carrying relevant biomedical information would be observed across cohorts. We decided to test this hypothesis by computing the similarity between pairs of SVN communities in different cohorts.

To quantify the similarity between pairs of SVN communities obtained for the different cohorts of patients, we compute the Jaccard similarity $J(C_{k, a},C_{\ell, b})$ between community $a$ of cohort $k$ $C_{k, a}$ and community $b$ of cohort $\ell$ $C_{\ell, b}$ as follows:
\begin{equation}
    J(C_{k, a},C_{\ell, b})=\frac{|\text{Edges}(C_{k, a}) \cap \text{Edges}(C_{\ell, b})|}{|\text{Edges}(C_{k, a}) \cup \text{Edges}(C_{\ell, b})|}
\end{equation}
where $\text{Edges}(C_{k, a})$ defines the set of links connecting the nodes of community $C_{k, a}$. The intersection between sets (labeled by the symbol `$\cap$') contains links present in both sets, whereas the union (`$\cup$') contains all links present in at least one of the two sets. The value of Jaccard similarity ranges from zero (when no link is present in both communities) to one (when both communities contain the exact same sets of links). 

Across all cohorts in Table~\ref{tab:comm_svn} we find a total of 380 communities with more than 25 nodes each. Across all pairs of communities among this set, the Jaccard similarity runs from 0 to the maximum value of 0.609. Computing the $380\times 380$ matrix of pairwise Jaccard similarities between communities reveals that a large fraction of community pairs have a similarity value close to 0.6, indicating a large overlap of edges present in several communities.

Starting from the Jaccard similarity, we define a distance for each pair of communities as $d(C_{k,a}, C_{l,b})=1-J(C_{k,a}, C_{l,b})$. Using this distance we perform an agglomerative hierarchical clustering procedure. Agglomerative Hierarchical Clustering is an unsupervised bottom-up clustering method where each element starts in its own cluster, and clusters are iteratively merged based on a decreasing similarity (i.e., increasing distance) measure until all objects form a single cluster. The clustering procedure starts with $n$ isolated elements, each containing a single element, computes the pairwise distance matrix between all observations and then proceeds with an iterative merging by find the closest pair of elements according to a linkage criterion (e.g., Single, Complete, Ward's or Average linkage). The sequence of merges is summarized in a hierarchical tree (also called dendrogram). In this study we use the average linkage algorithm \cite{kaufman2009finding}. This leads to the tree shown in Fig. \ref{fig:HTcolor}. For future purposes each of the 380 communities is assigned a running number (1 to 380) in the order as they appear in the tree.

We find that the hierarchical tree is quite structured, indicating that the SVNs have communities of different patient cohorts that cluster together. In other words, there are similarities among communities of diseases that persist over cohorts of different age intervals and different sexes. The coherence of groups of communities detected by hierarchical clustering in terms of their membership to specific cohorts and categories of diseases can be seen in detail by analyzing the Jaccard similarity matrix in Fig. \ref{fig:HT_Jaccard} in Appendix \ref{appendix_jaccmat}. The rows and columns of the matrix are arranged in the order as they appear in the tree in Fig. \ref{fig:HT_Jaccard} to efficiently visualize the clustering of groups of communities.

\subsubsection{Over-expression of ICD codes in different communities}
\label{sec:community-over-exp}
\begin{sidewaysfigure*}
\centering
    \includegraphics[width=\columnwidth]{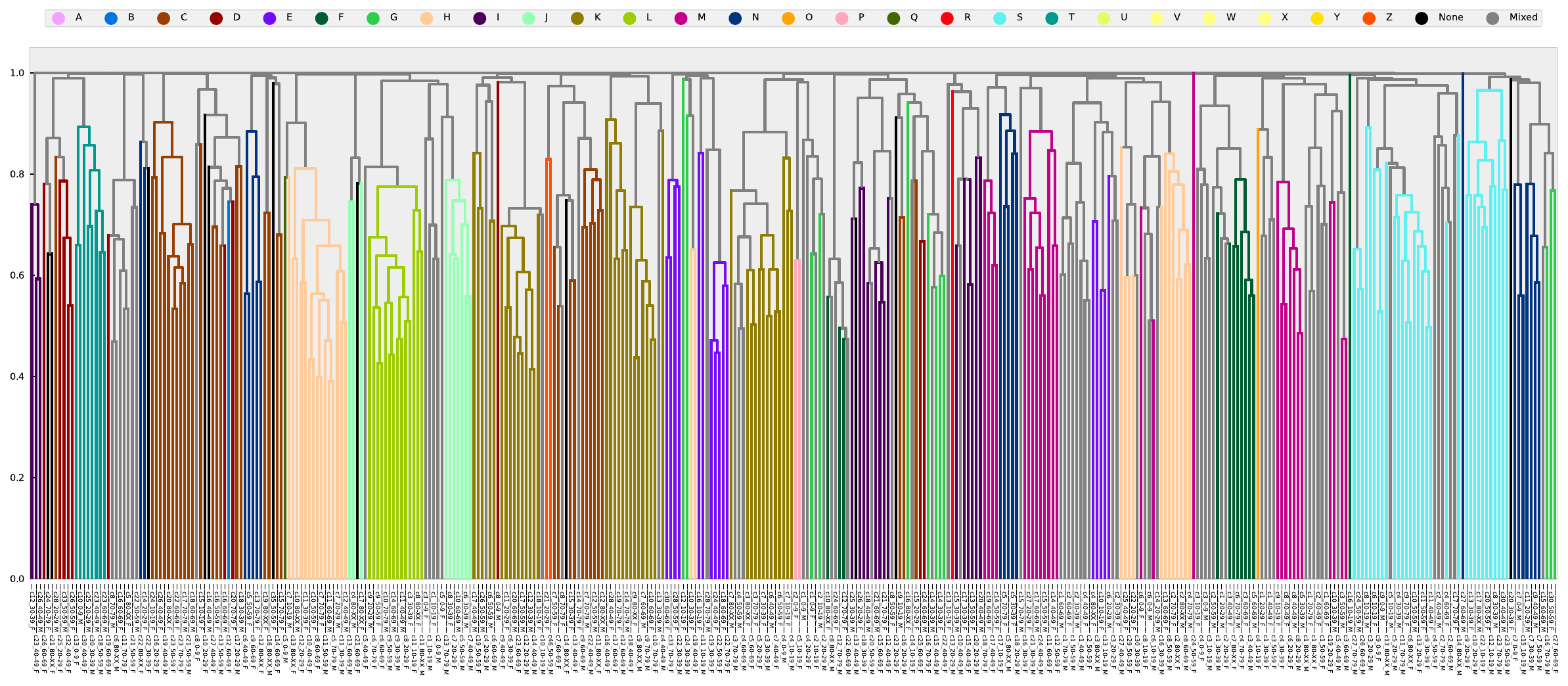}
\caption{Average linkage hierarchical tree of the 380 ICD communities with more than 25 nodes detected in SVNs of different cohorts. The average linkage tree is obtained by using $d=1-J$ as a dissimilarity measure. The color of each leaf of the tree indicates over-expression of an ICD category as reported in the color spots shown above the tree. If two leaves of the same color are joined together, the corresponding branch will have the same color. Otherwise, the branch will be grey. When we observe over-expression of more ICD categories, we draw the line corresponding to the community as a gray line. Communities without over-expression of ICD categories are drawn in black lines.}
\label{fig:HTcolor}
\end{sidewaysfigure*}

We find that different communities are characterized by an abundance of ICD codes of one or  a few categories. To confirm this observation quantitatively, we perform a statistical test evaluating the over-expression of ICD codes of a specific category for each community. The statistical test is analogous to the one performed to detect statistically validated networks (details can be found in reference \cite{tumminello2011community}). Based on the statistical test, we conclude that 369 communities out of 380 present over-expression of one or more ICD categories. In Fig. \ref{fig:HTcolor}, when we observe an over-expression of ICD codes with just a single specific letter in a specific community, we draw the given community with a color associated with the over-expressed category. 

Large branches of the same color are noticed in Fig. \ref{fig:HTcolor}, indicating that these sets of communities are rich in a specific category of diseases. An overall analysis of all over-expression shows that 293 communities present over-expression of just one category, 70 communities of two categories, 13 communities of three categories, and one community has over-expression of 6 categories. The high number of communities with over-expression of one (or more than one) ICD category suggests that the partition in communities of SVNs highlights information that can be interpreted in terms of specific groups of diseases.

\begin{figure*}
  \centering
    \includegraphics[width=\linewidth, trim=0 0 0 60, clip]{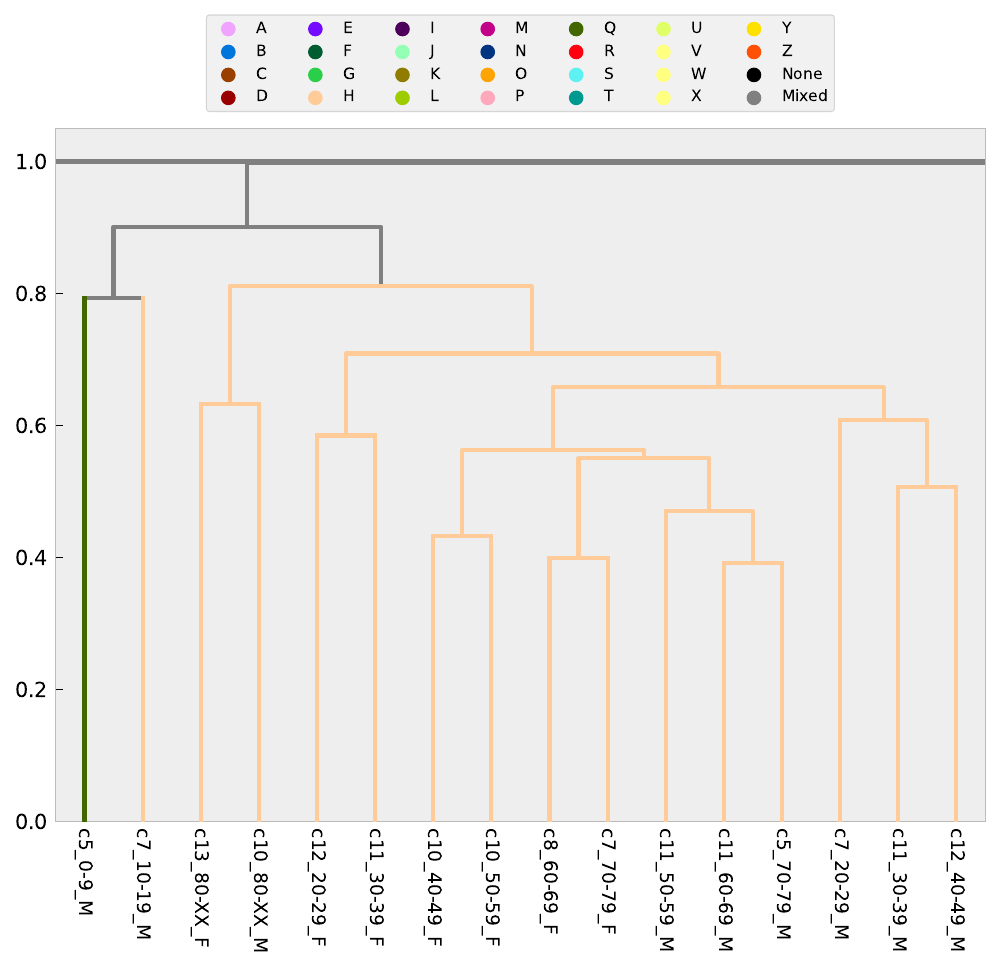}\\
  \caption{Selected region of the average linkage hierarchical tree of the set of SVN communities with number of nodes larger than 25. The selected region contains the pair of ICD communities with the highest Jaccard similarity (c5\_70-79\_M and c11\_60-69\_M). All communities except one (c5\_0-9\_M) in this cluster present an over-expression of the ICD category H.}
  \label{fig:HT_AL_65_80}
\end{figure*}

 We label different communities by codes such as  c11\_60.69\_M and c5\_70.79\_M For example. The former is the community with numeric label 11 of the cohort of ages 60-69 for male patients,   and the latter is the community with numeric label 5 of the cohort of ages 70-79 again for males. These two communities are located in the portion of hierarchical tree shown in Fig. \ref{fig:HT_AL_65_80}. The selected branch of hierarchical tree shows that the clustering of communities involves cohorts of both sex and of different age intervals. This branch of communities shows an over-expression if ICD codes in category H, and with numerical codes ranging from 60 to 90 therefore referring to diseases involving the ear and hearing. 

The presence of the statistically validated over-expression suggests a rich pattern of comorbidity involving diseases of different types and/or affecting different organs or physiological districts. The complete list of over-expressions of ICD communities is provided in the tables in Appendix \ref{appendix_oe25}, where we detect long sequences of communities (when ordered as they appear in the hierachical tree) with over-expression of the same ICD category. The high level of homogeneity of over-expression of the same ICD category in subregions of the hierarchical tree shows that detected clusters are, in most cases, associated with a specific category of diseases. The list of over-expression categories also highlights the presence of a few large clusters characterized by over-expression of more than one ICD category. In summary, ICD communities detected in SVN comorbidity networks partition the comorbidity space and provide robust information on the degree of similarity or differences between each pair of ICD communities specific for age interval and sex.


\subsection{Analysis of community clusters: The case of mental and behavioral disorders} \label{sect:case_study}

\begin{figure*}
\centering
  \centering
    \includegraphics[width=\linewidth, trim=0 0 0 60, clip]{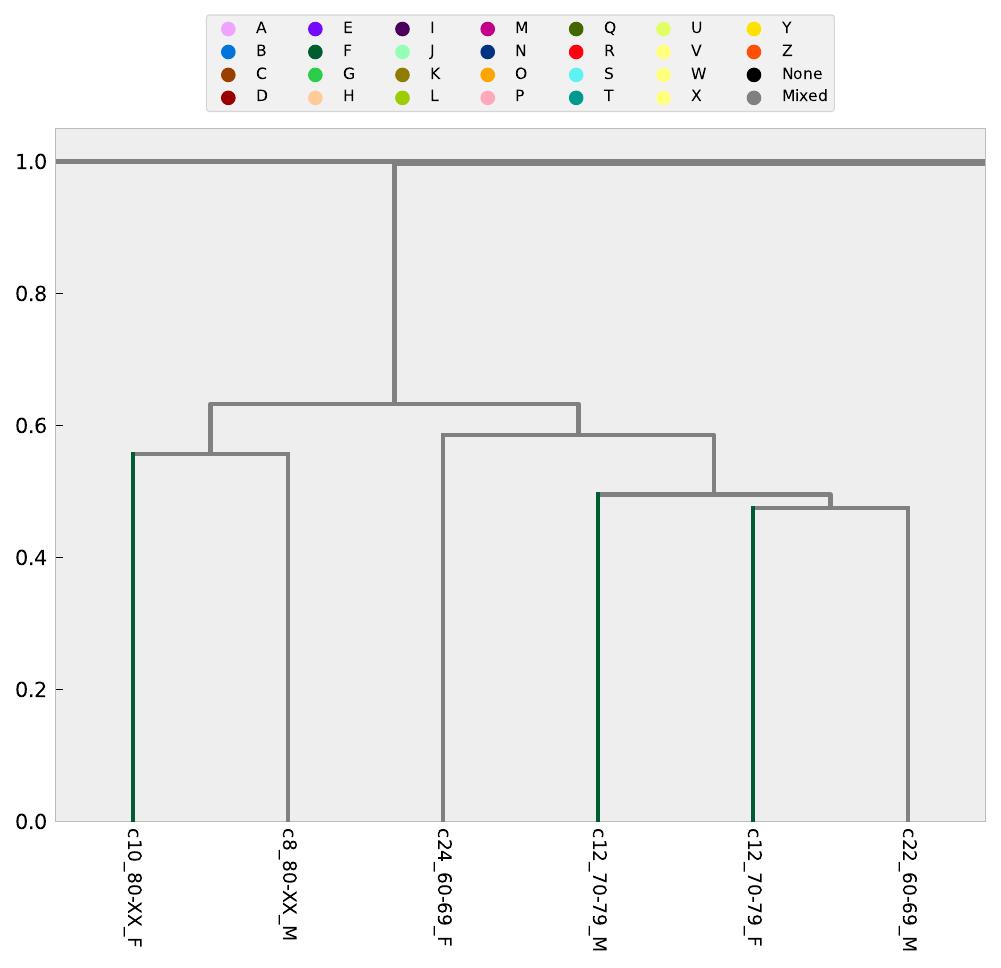}
  \caption{Selected region of the average linkage hierarchical tree with communities located at positions from 199 to 204. The over-expression of disease category for each community starting from left is as follows F,~FG,~FG,~F,~F,~FG.}
  \label{fig:HTAL_sub1}
\end{figure*}

\begin{figure*}
  \centering
    \includegraphics[width=\linewidth, trim=0 0 0 60, clip]{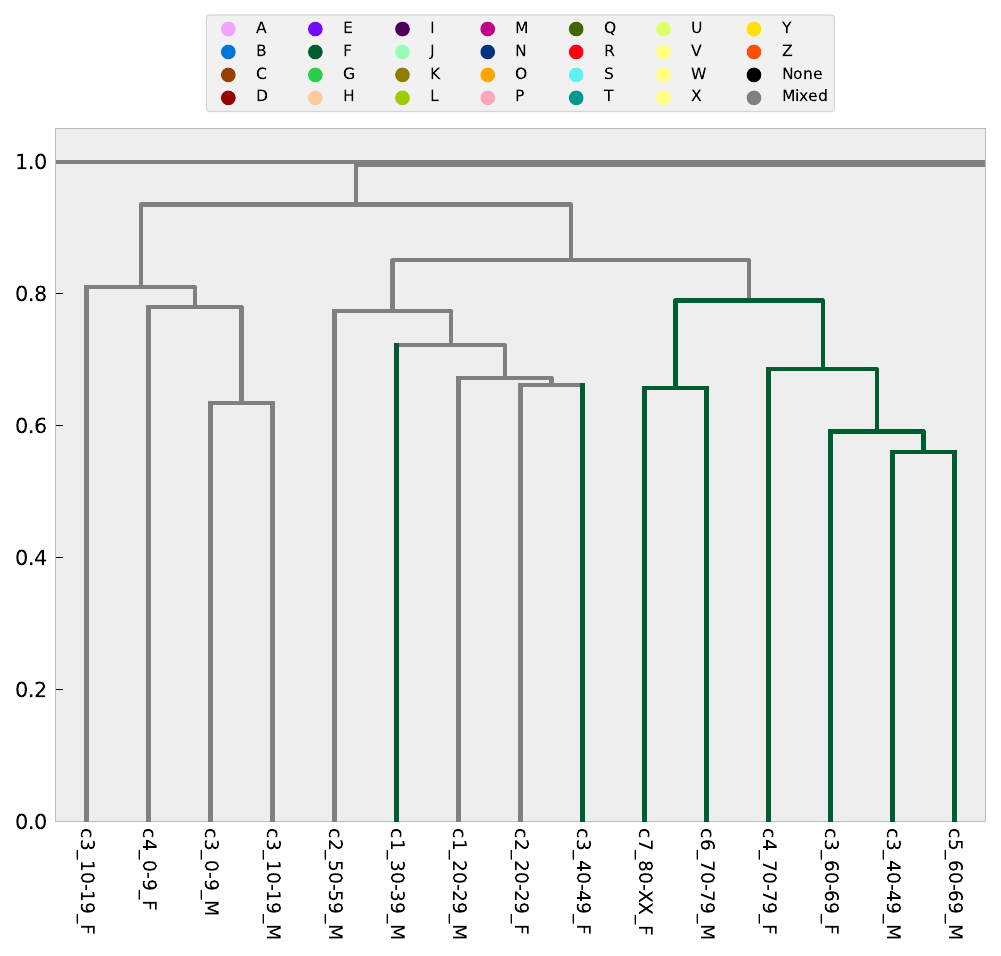}
  \caption{Selected region of the average linkage hierarchical tree with communities located at positions from 291 to 305. The over-expression of categories for each community, starting from left, is as follows FYZ,~FZ,~FZ,~FZ,~FS,~F,~FZ,~FZ,~F,~F,~F,~F,~F,~F,~F.}
  \label{fig:HTAL_sub2}
\end{figure*}

In this subsection, we provide two examples of analyses of clusters of communities detected in the SVNs and characterized by the over-expression of a specific ICD category. We discuss clusters of communities characterized by the over-expression of category F (mental and behavioral disorders). Two regions of the average linkage hierarchical tree present this over-expression (see complete list of category over-expression in Appendix~\ref{appendix_oe25}). They are the cluster comprising communities located in the average linkage hierarchical tree from position 199 to 204 (see Fig. \ref{fig:HTAL_sub1}) and the cluster of communities located from position 291 to 305 (see Fig. \ref{fig:HTAL_sub2}). 


\begin{figure*}
   \hspace*{-0.5cm}\includegraphics[width=\linewidth]{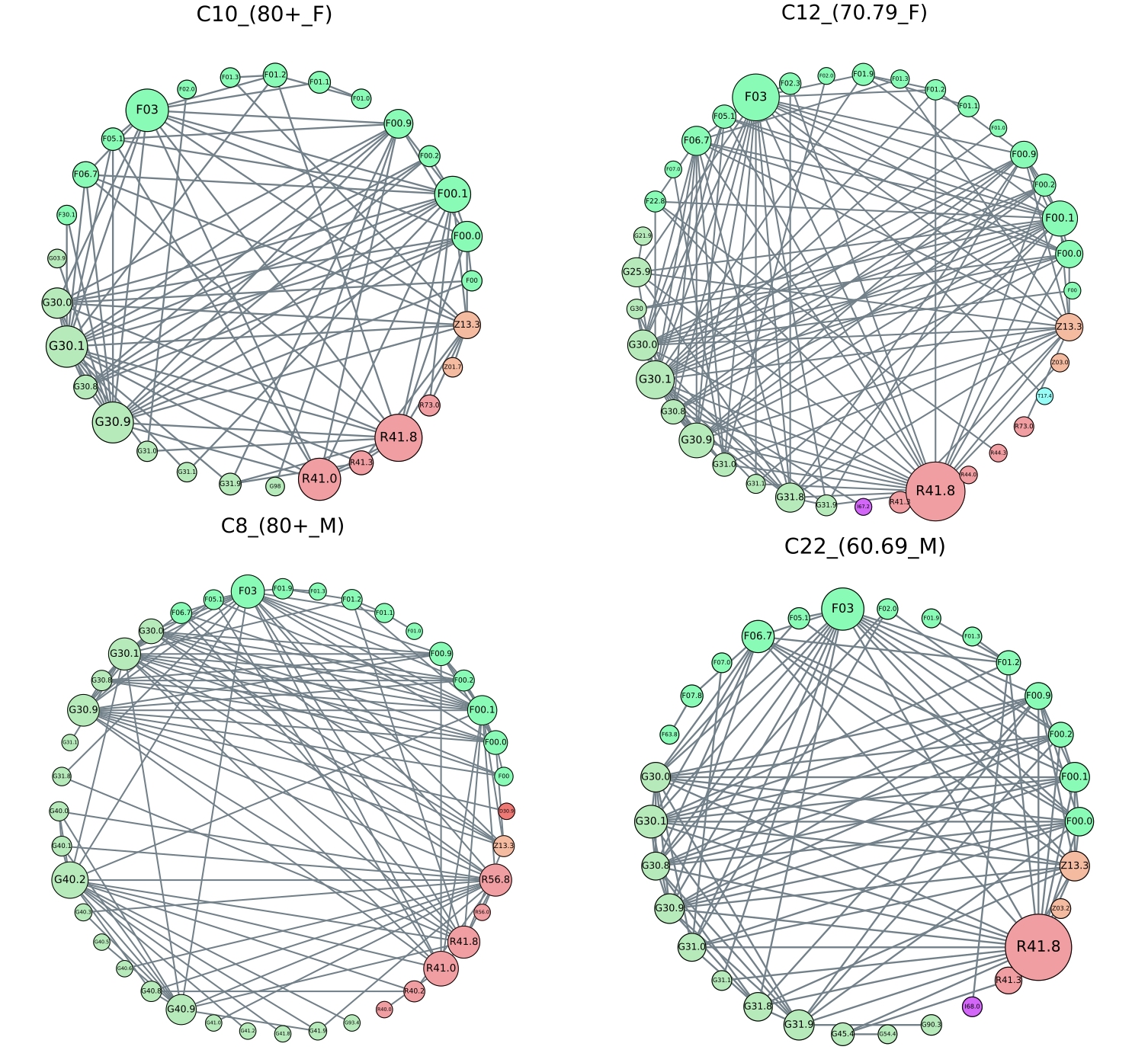}
  \hspace*{-0.5cm}\caption{Four communities c10\_80.XX\_F, c12\_70.79\_F, c8\_80.XX\_M, and c22\_60.69\_M of the community cluster shown in Fig. \ref{fig:HTAL_sub1}. Different node colors indicate different ICD categories. The size of each node is proportional to the degree of the node in the SVN.}
  \label{fig:F_c1_comm}
\end{figure*}

The first group (Fig.~\ref{fig:HTAL_sub1}) consists of 6 communities of ICD codes from cohorts with age intervals from 60-69 to 80+ for females and males. Four of the communities in this cluster are shown in Fig.~\ref{fig:F_c1_comm}. In addition to the F over-expression in three communities, we also observe over-expression of the G ICD category (diseases of the nervous system). The majority of the diseases present in this cluster concerns degenerative diseases of the nervous system coded in the set of ICD codes ranging from G30.* to G39.*. 

\begin{figure*}
  \centering
    \includegraphics[width=\linewidth]{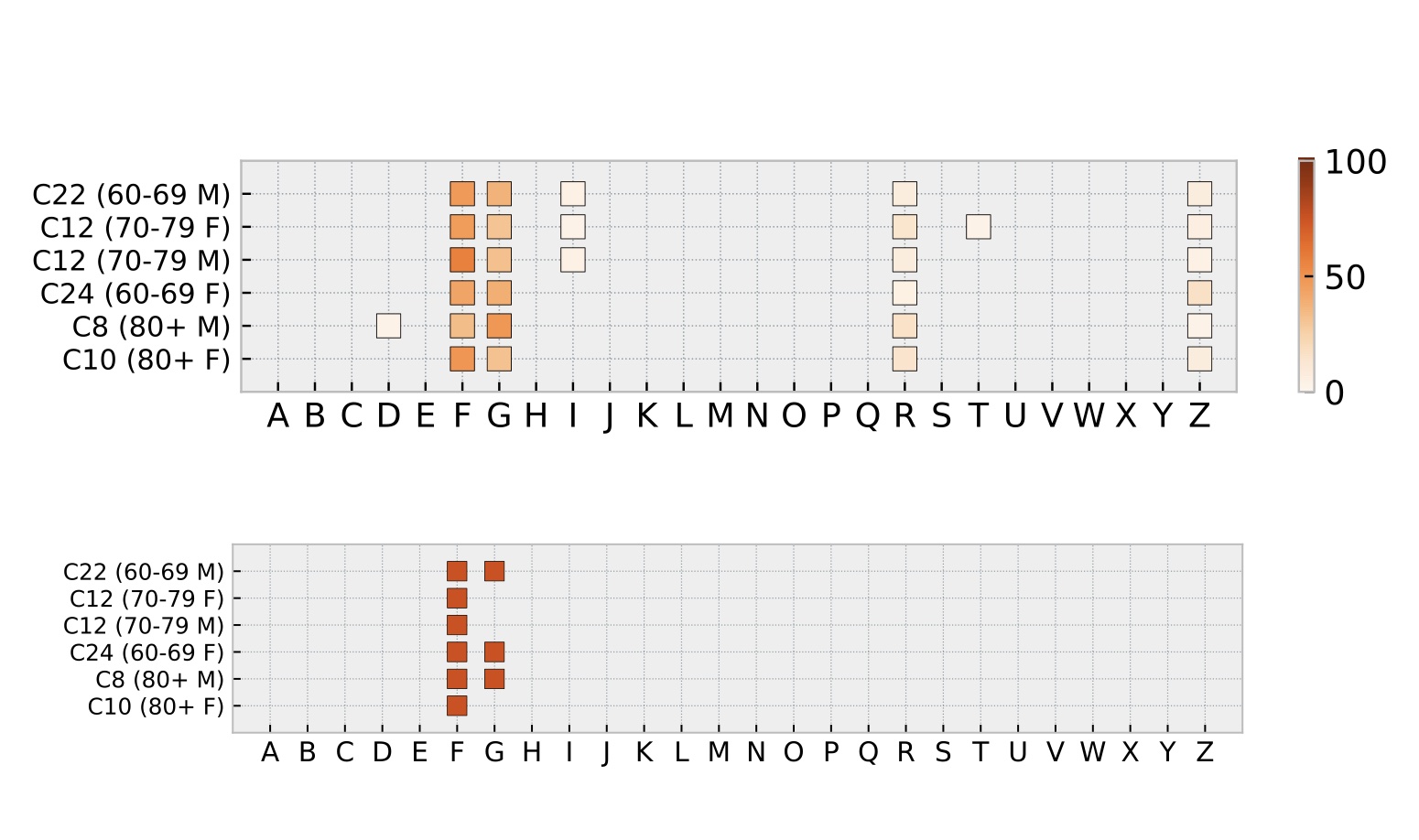}
  \caption{Occurrence (upper panel) and over-expression (lower panel) of the different ICD letter categories in the communities in the cluster of Fig.~\ref{fig:HTAL_sub1}. The color scale in the upper graph represents the percentage of nodes belonging to a particular disease class (letter) over all the classes. In the lower graph, the color is fixed since over-expressions are either present or not.}
  \label{fig:F_c1_OccOE}
\end{figure*}

In Fig. \ref{fig:F_c1_OccOE}, we show information about the occurrence (top panel) and the over-expression (bottom panel) of ICD codes present in each disease's community of the cluster. We note that the occurrence of ICD codes is almost exclusive to the categories F, G, R, and Z and a few additional diseases of I, D, and T categories. This cluster is therefore specific for degenerative disease of the nervous system affecting both females and males starting from the age of sixties. ICD codes R concern primarily symptoms, and ICD codes Z primarily describe anamnesis and/or specific health procedures or circumstances (see Table \ref{tab:letters} for the full definitions of all categories). The most prominent ICD code indicating symptoms we note is R41.8 (``other and unspecified symptoms and signs involving cognitive functions and awareness"), i.e., the most basic evaluation accompanying degenerative diseases of the nervous system. For the community c8\_80.XX\_M we also note ``Other and unspecified convulsions" that in this community is connected to several diseases of the group G40.* used for epilepsy.

The second group of ICD communities (see Fig. \ref{fig:HTAL_sub2}) consists of three distinct subclusters. These subclusters are different in terms of age intervals and prominence of certain types of mental and behavioral disorders. The subcluster composed of c3\_10.19\_F, c4\_0.9\_F, c3\_0.9\_M, and c3\_10.19\_M (first branch on the left region of Fig. \ref{fig:HTAL_sub2}) covers the age cohorts of childhood and adolescence and, in fact, the ICD codes are primarily in the range F9*. This group of codes corresponds to ``behavioral and emotional disorders with onset occurring in childhood and adolescence". In addition to the prominent F and Z categories, these communities also show some ICD codes of categories P (certain conditions originating in the perinatal period) and Q (congenital malformations deformations and chromosomal abnormalities).

\begin{figure}[h]
  \centering
    \includegraphics[width=\linewidth]{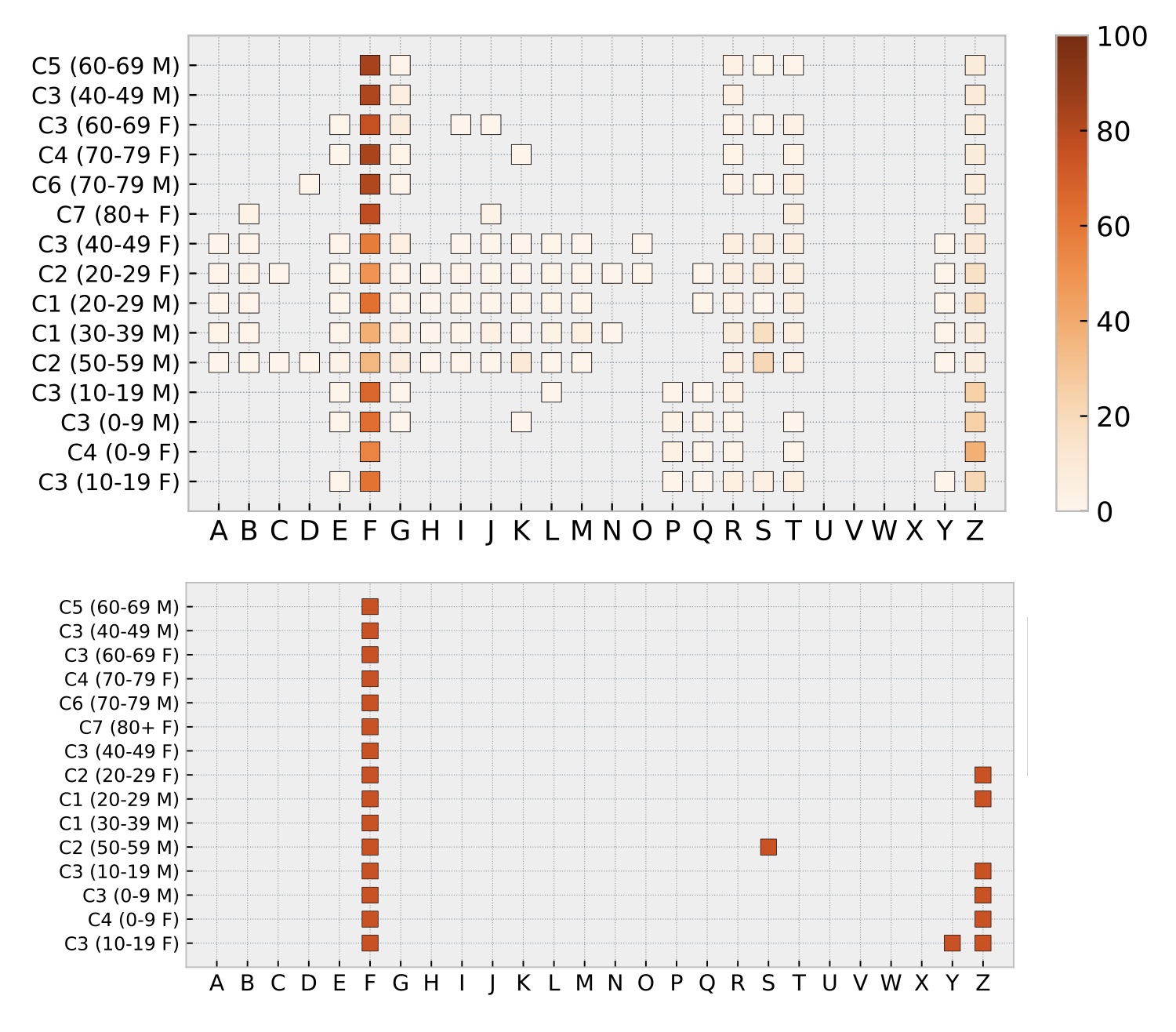}
  \caption{Occurrence and over-expression of ICD letter category in the communities in the cluster of~Fig. \ref{fig:HTAL_sub2}.}
\label{fig:F_c2}
\end{figure}

The second branch (middle branch in Fig. \ref{fig:HTAL_sub2})  covers the age period from twenty to sixty. It comprises the five communities c2\_50.59\_M, c1\_30.39\_M, c1\_20.29\_M, c2\_20.29\_F, and c4\_40.49\_F. These are communities with a large number of nodes and links, including some nodes belonging to many different disease categories (see top panel of Fig. \ref{fig:F_c2}). However, in spite of the presence of nodes of several different categories, the over-expressed categories are F and Z, showing that the backbone of these communities consists of mental and behavioral disorders. 


The third branch (right branch in Fig. \ref{fig:HTAL_sub2})  has primarily communities of older patients. The six communities in this branch are c7\_80.XX\_F, c6\_70.79\_M, c4\_70.79\_F, c3\_60.69\_F, c3\_40.49\_M, and c5\_60.69\_F. These communities contain nodes from multiple categories of ICD codes (see top panel of Fig. \ref{fig:F_c2}). Specifically, in addition to the over-expressed F category, we observe the G, R, T, and Z categories. The presence of the G category reflects the role of diseases of the nervous system whereas the S and T categories represent ``injury, poisoning and certain other consequences of external causes". For this subcluster, the only over-expressed category is F. For these age cohorts, the differences between males and females decrease when age increases.

In summary, the comparative analysis of ICD code communities obtained from statistically validated comorbidity networks gives us interesting insights about how disease classes are distributed across different cohorts of patients of different ages and sexes. Here, we have presented the case of disease communities with an over-expressed presence of diseases belonging to the category of mental and behavioral disorders (F class) but our results cover all ICD categories, and a similar detailed analysis can be performed for any category of interest.

For example, when we consider communities involving diseases primarily related to categories C and D (i.e., Neoplasms) we note that the node of highest degree of each community is informative with respect to the branch of hierarchical tree the community belongs to. In fact, for each cluster observed in the hierarchical tree with over-expression of category C, communities of a given branch of the hierarchical tree are rich in information about comorbidity associated with a specific type of neoplasm. Below, we report the ICD node of the highest degree that is present in each branch with C over-expression. Specifically, at different positions of the hierarchical tree, we observe: for positions 4-11 (D48 - Neoplasm of uncertain or unknown behavior of other and unspecified sites), for 20-27 (C43/44 - Malignant melanoma of skin/Other malignant neoplasms of skin), for 31-41 (C83 - Non-follicular lymphoma), 42-53 (C90 - Multiple myeloma and malignant plasma cell neoplasms), for 61-63 (C56 - Malignant neoplasm of ovary), 131-136 (C50 - Malignant neoplasm of breast), for 137-143 (C79 - Secondary malignant neoplasm of other and unspecified sites), and for 216-228 (C71 - Malignant neoplasm of brain).

\subsection{Dismantling the comorbidity network} \label{sect:dismantling}

A classical topic in network theory is the investigation of the level of network resilience to random or targeted attacks \cite{Albert2000}. This is relevant under two different view points: on one side such levels of resilience are indicative of how much networks are robust against attacks. On the other side, however, looking for the best strategy to dismantle a network can give an indication of what are the nodes or group of nodes most relevant in guarentiing the network cohesivity. In the present context, identifying such peculiar nodes or groups of nodes, might give us information on the  pathologies or sets of pathologies that are central in the setting of comorbidities. 

\subsubsection{The dismantling procedure}
The presence of a link in the comorbidity network is carrying two types of information. On one hand it is showing that the two diseases occurs in a patient with a probability that it is not compatible with a null hypothesis based on the prevalence of the two diseases. The same link is also signaling the potential for patients having only one of these diseases at a given time to get the second one. This means that developping comormidities over time for a specific patient can be seen  as a walk performed on the comorbidity network. Imagine a disease path characterized by diseases A, B and C. Dismantling the comorbidity network (for example due to preventive medicine policies affecting disease B) therefore would diminish groups of patients affected by disease A to develop comorbidities with disease C (and viceversa) thanks to the hipothesized dismantling action.

Let us now consider a simulated ``dismantling" procedure \cite{wandelt2018comparative} of the PROJ and SVN comorbidity networks to highlight the categories of ICD codes which are most important for bridging different regions of the comorbidity network.  For each age and sex cohort, we remove one node (i.e., one ICD code) at a time from the largest connected component of the comorbidity network, iterating the removal process until percolation across all remaining nodes of the comorbidity network breaks down. We monitor the presence of percolation among the remaining nodes of the largest connected component by comparing the size of the first and second largest connected components at each removal step. In the presence of percolation, the largest connected component is much bigger than the second largest connected component. In contrast, when percolation is lost, the first and the second largest connected components have comparable sizes \cite{musciotto2023exploring}. 

The sequence of node removal of ICD codes is obtained using the following procedure: at each step, we compute the betweenness of all nodes, and we remove the node with the highest betweenness \cite{wandelt2018comparative}. The dismantling process is applied to a slightly modified version of the comorbidity networks discussed in Sect. \ref{sec:svn-props}. Specifically, for each cohort the  modification we make consists in removing all nodes with ICD codes belonging to categories R and Z. These ICD categories either represent codes for symptoms  or are used by physicians to document the type of interactions patients had with the health services. These nodes connect diseases through common symptoms or hospital procedures/check-ups. Thus, removing these nodes from the network has a different meaning than removing a node representing an actual disease. In this section, we want to focus on how ``eradicating" a disease can alter the structure of the PROJ network or SVN, and how important each one of them is in holding the largest connected component together. 

\subsubsection{An illustrative example}
\begin{figure*}
\centering
    \includegraphics[width=\linewidth, trim=0 100 0 0, clip]{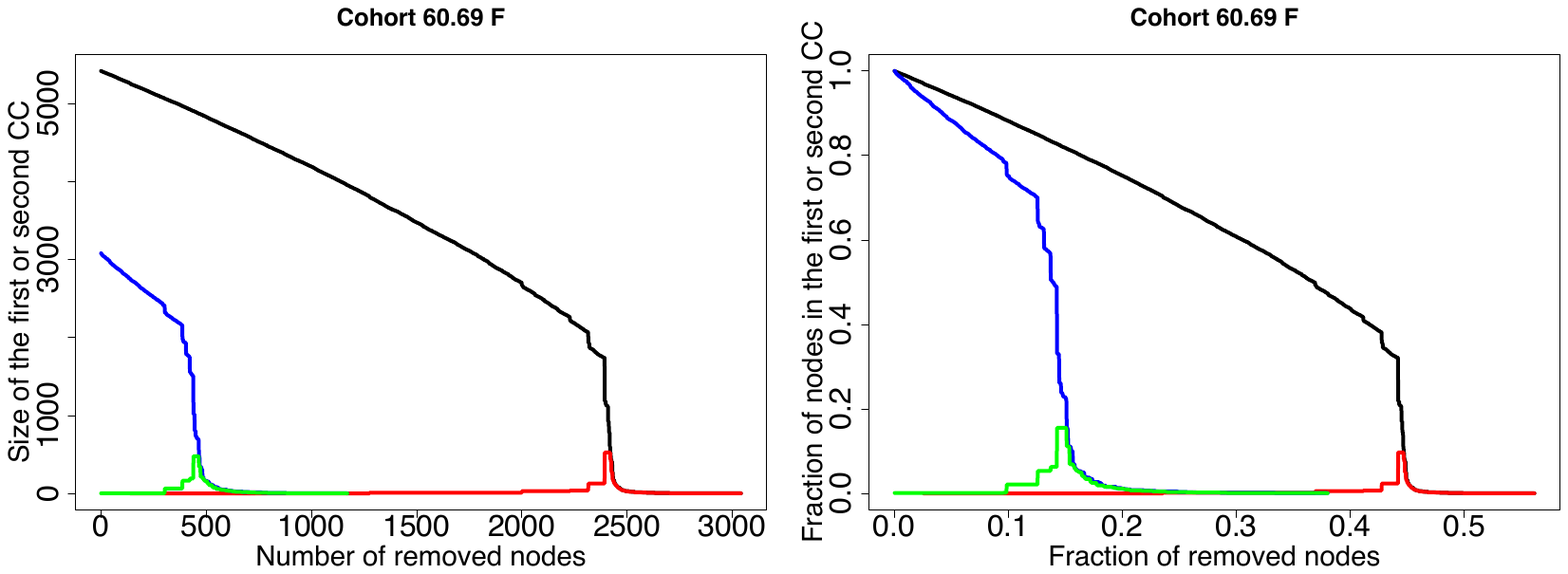}
\caption{Dismantling process of the comorbidity networks obtained for female 60-69 cohort. In the two panels, we show the size of the largest connected component (black line for PROJ network and blue line for SVN) and the size of the second largest connected component (red line for the PROJ network and green line for the SVN) as a function of the number of removed nodes. The left panel shows the dismantling process in absolute terms, whereas the right panel in terms that are relative to the size of the LCC.}
\label{fig:dis1}
\end{figure*}

In Fig.~\ref{fig:dis1}, we show results for the dismantling process for a specific cohort of patients (female, aged 60-69). More specifically, we plot the sizes of the largest and second largest connected components as the removal process unfolds. In the plot, we illustrate the dismantling process of the PROJ network (black and red lines) and of the corresponding SVN (blue and green lines). We find that the dismantling procedure of the SVN is more efficient than that for the PROJ network, in the sense that it requires the removal of only about 15\% of the nodes of the SVN largest component (which contains about 57\% of the nodes of the corresponding projected network). Indeed, for all cohorts, we verify that the dismantling of the SVN is more efficient than the dismantling of the PROJ network both in absolute and relative terms. The right-hand panel of Fig.~\ref{fig:dis1} shows that the fraction of nodes that has to be removed from the PROJ network to cause its collapse is more than double than the fraction that needs to be removed from the SVN.

\subsubsection{Categories of dismantling nodes}
We now investigate the type of ICD codes that contribute most to the dismantling of the comorbidity network of a given cohort.  For each cohort, we first select a set of nodes comprising the first $N_{\text dism}$ ICD codes whose removal (as a whole) reduce the size of the largest connected component of the SVN to 10\% of its original size. We call this the `dismantling set'. The number $N_{\text dism}$ will be specific to each cohort. For each disease category, we calculate the fraction of nodes that are in the dismantling set and we compare this fraction to the fraction observed by randomly sampling a 1000 sets of ICD nodes of the same size of the dismantling set from the corresponding SVN network. This comparison allows us to detect the relative abundance of nodes of a specific category in the dismantling set for each cohort.

\begin{figure*}
\centering
    \includegraphics[width=\linewidth]{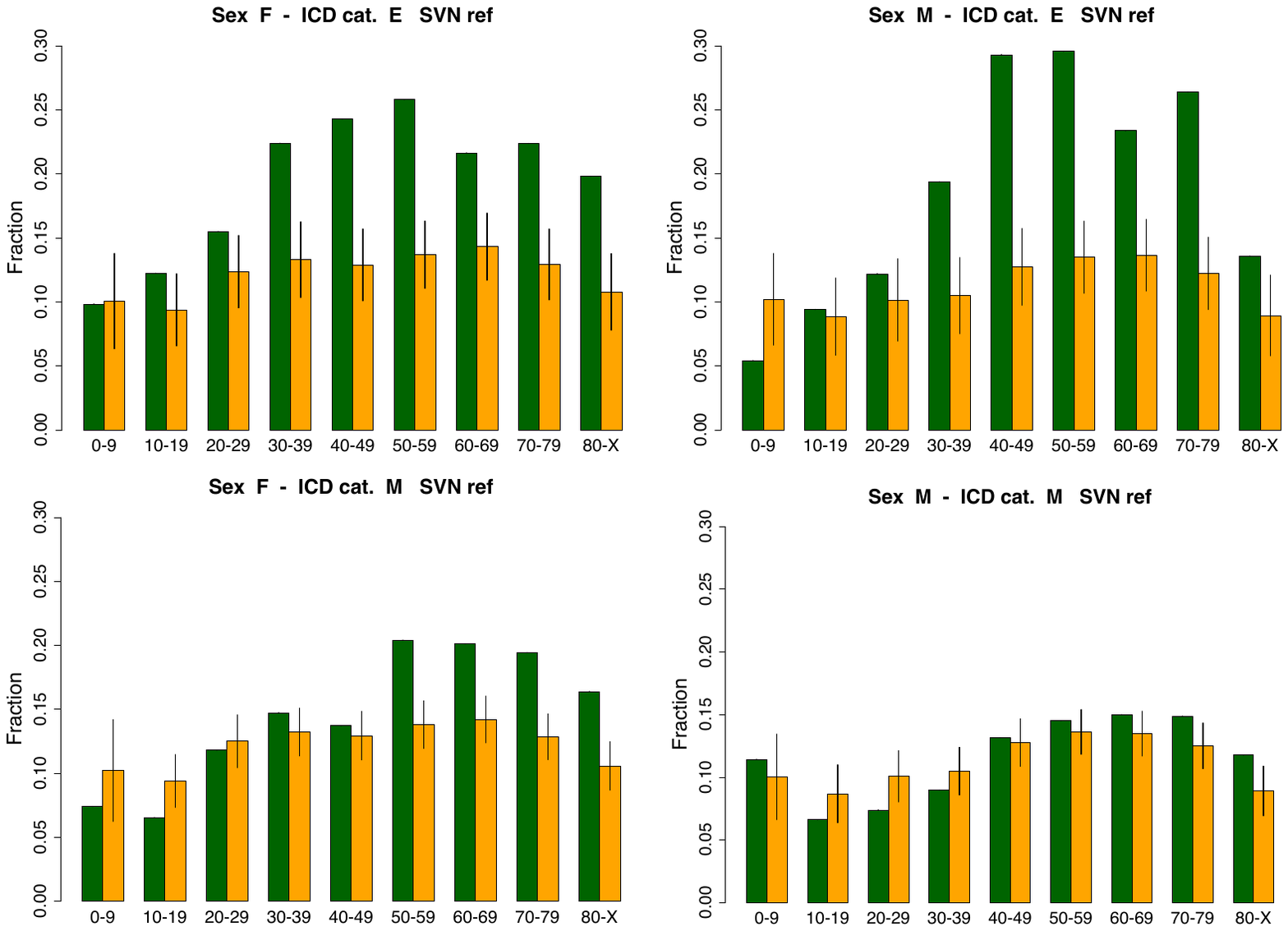}
\caption{Top panels: Fraction (dark green bars) of removed ICD codes reducing the size of the largest connected component of SVN to 10\% of its original size for the category E: Endocrine, nutritional and metabolic diseases (top panels), 
and for the category M: Diseases of the musculoskeletal system and connective tissue (bottom panels), for all age cohorts of females (left panel) and males (right panel). Orange bars are the corresponding fraction of category E obtained by random selection of nodes of the SVN networks, obtained by selecting a number of nodes equal to the size of "dismantling" nodes detected for each cohort. Black segments on top of the orange bars display plus and minus one standard deviation observed across 1000 realizations of the sampling procedure. Bottom panels: the same quantities for category M (Diseases of the musculoskeletal system and connective tissue).}  
\label{fig:bar1}
\end{figure*}

In Fig.~\ref{fig:bar1} we show two examples of the frequencies of ICD codes in different categories among dismantling sets in the different age cohorts for females (left-hand panels) and males (right-hand panels). Specifically, we focus on categories E (endocrine, nutritional and metabolic diseases, top panels) and M (diseases of the musculoskeletal system and connective tissue, bottom panels). Dark green bars in the figure represent the frequency observed in the dismantling set for the cohort in hand, and orange bars the average frequency of 1000 simulations of random selection from the corresponding SVN network. Vertical black lines on top of the orange bars indicates plus and minus one standard deviation across the 1000 samples.

The two examples in Fig.~\ref{fig:bar1} show two typical patterns. Specifically, the case of category E (upper panels) shows a frequency pattern for different age cohorts that is quite similar for both sexes. The frequency of the E category of diseases is compatible, or almost compatible, with the one expected for a random inclusion in the dismantling set for age cohorts 0-9, 10-19, 20-29 and 80-XX. On the other hand, a higher frequency is observed for the age cohorts 30-39, 40-49, 50-59, 60-69 and 70-79. In summary, in the age period from 30 to 80 years old, endocrine, nutritional, and metabolic diseases have a localization in the comorbidity networks that gave them high values of node betweenness therefore putting these diseases on a large number of diseases paths observed in the comorbidity networks.

A quite different pattern is observed for diseases of category M, see the lower panels in Fig.~\ref{fig:bar1}. These are conditions related to the musculoskeletal system. For male patients, the frequency of category-M ICD codes in the dismantling set is consistent with the one estimated from the random sampling from the SVN network. In other words, for males, we verify that category-M diseases do not sustain resilience of the comorbidity network relative to the basic level expected by random co-occurrence. However, this observation is only valid for males. In fact, for females, we detect a different pattern. Absence of enhancement of frequency of M diseaases is observed for younger age cohorts (until age 49) but an abrupt frequency enhancement is observed starting from the 50-59 age group. For higher ages, the enhancement is observed for all female cohorts.
\begin{figure*}
\centering
    \includegraphics[width=\linewidth, trim=0 100 0 0, clip]{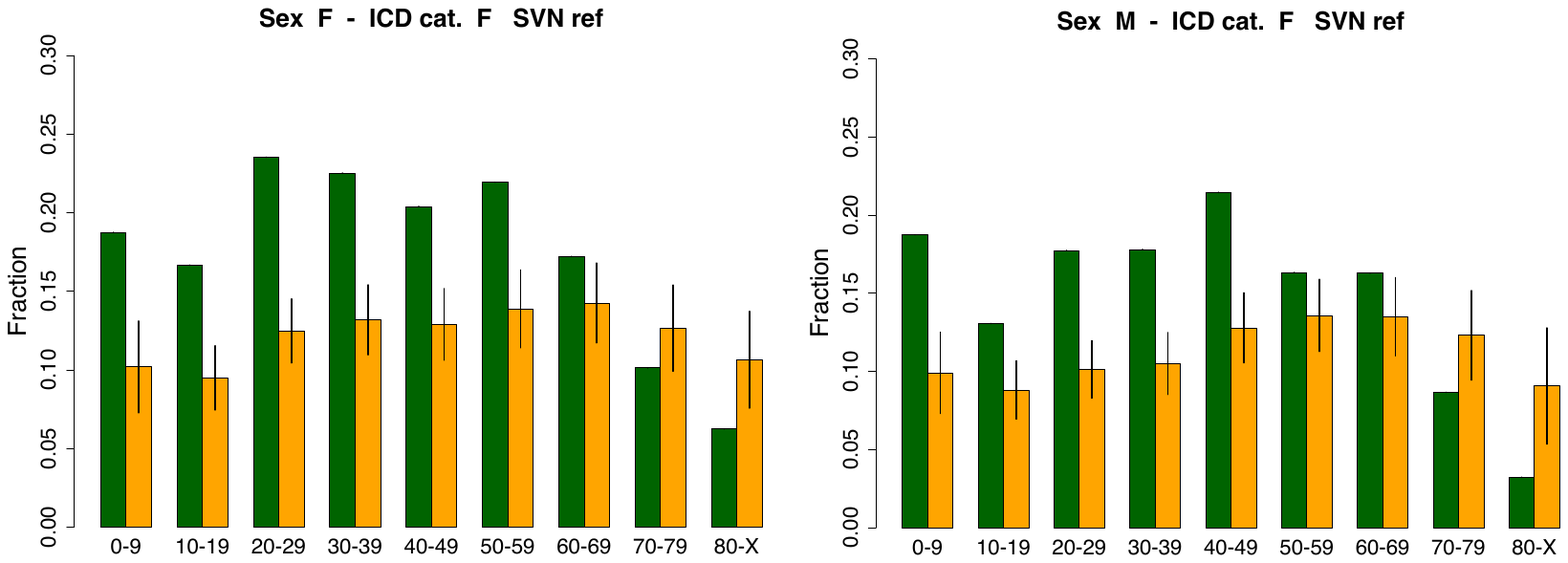}
\caption{Fraction (dark green bars) of removed ICD codes reducing the size of the largest connected component of SVN to 10\% of its original size for the category F (Mental and behavioural disorders) for all age cohorts of females (left panel) and males (right panel). Orange bars are the corresponding fraction of category F obtained by randomly selecting in the SVNs a number of nodes equal to the size of "dismantling"  nodes detected for each cohort. Black segments on top of the orange bars display plus and minus one standard deviation observed in 1000 random realizations.}  
\label{fig:barF}
\end{figure*}

As a last example of a disease category with enhanced frequency in the dismantling set of nodes across different cohorts, we comment on ICD category F. These are the mental and behavioral disorders we also discussed in Sec.~ \ref{sect:case_study}. In Fig.~\ref{fig:barF}, we notice that F ICD nodes show a higher frequency in dismantling nodes of younger cohorts both for females and males compared to random sampling. The presence of these codes is enhanced up to cohorts of age 60 to 69 in females and 50-59 in males. There are different explanation for the key role of diseases in the F group for cohorts involving children and teens (cohorts of 0-9 and 10 to 19 of age) and cohorts involving young adults (cohorts of 20-29 of age), respectively. In the former cohorts, the ICD codes  of type Finvolve diseases associated with mental retardation, whereas starting from late teens to young adults different forms of depression play a major role. The female (left-hand panel) and male (right-hand panel) patterns are similar but not identical across different age cohorts. This suggests a certain degree of sex-specificity associated with different impacts of mental retardation and different degree of impact/resilience with respect to depression.

For each disease category, the dismantling method is able to highlight the classes of diseases that have a localization in the comorbidity networks that gave them high values of node betweenness therefore shortening the paths of the comorbidity network, providing suggestions for preventive medicine policies.

\section{Discussion and conclusions} \label{sect:conclusions}

In this paper, we investigate comorbidity patterns observed in a large set of patients diagnosed in southern Finland in a time period of 15 years. Evidence of comorbidity patterns is obtained by using information about diseases diagnoses that are present in the historical medical records of patients of a large population. Starting from a bipartite network of patients and diseases we construct comorbidity networks of diseases where a link indicates an over-expression of comorbidity that is statistically validated against a null hypothesis of random co-occurrence of a pair of diseases in patients of a given age and sex. 

Our study is performed at so-called level 4 of the ICD WHO code. It is worth to recall that this level is the one primarily used in medical diagnoses. Therefore, this level of granularity is probably the one that better describe the information stored in diagnoses by medical doctors. 
The redundant or inessential information associated with this high level of granularity is taken into account by choosing a statistical validation methodology that is robust with respect to the heterogeneity of the nodes (e.g., with respect to the different prevalence of diseases) and by performing a careful control of the family-wise error rate.

We have verified that comorbidity SVNs show information that is hidden in basic projected disease networks (the networks that we called `PROJ'). SVN allows unsupervised detection of groups of diseases (labeled by us as `diseases communities'). ICD communities are obtained by using community detection algorithms introduced in complex network research. Disease communities  obtained in this way are rich of medical information that can be used as supporting information for healthcare policy decisions. For example, mental and behavioral disorders have a strong impact in comorbidity of cohorts of young patients of both sexes. Conversely, endocrine, nutritional, and metabolic diseases present a more prominent role in sustaining comorbidity pattern in age cohorts ranging from the 30s to the late 70s with a similar impact for both sexes. 

SVNs are also informative with respect to the category (or categories) of diseases that need attention to fragment comorbidity SVNs for different cohorts of age and sex. In fact, the nature and structure of SVNs vary for different cohorts and the relative role of different categories of diseases can be comparatively assessed.   

A prominent example of different behavior is observed for the diseases of the musculosketal system and connective tissue where a distinct impact is seen for females of age from 50s to late 70s whereas no apparent role is detectable for males at any age.

In summary, information stored in electronic health records of a large population of patients, whose history is recorded for many years, successfully contributes to highlighting the role of specific categories of diseases and of specific prominent diseases in the setting of complex comorbidity patterns observed in a large cohort of patients of specific age and sex.


\section{Declarations} \label{sect:declarations}

\subsection{Availability of data and materials} 

The data investigated in this study are proprietary data of Auria Clinical Informatics which operates in connection with Varha. Data can be accessed with permission from Varha. The present study analyzes disease networks obtained with the approval of the Institutional Review Board of Turku University Hospital (license number T152/2017 [35]). Informed consent was waived due to the study's retrospective design, according to Finnish legislation on the secondary use of health data.

\subsection{Competing interests}

The authors have no competing interests as defined by Springer, or other interests that might be perceived to influence the results and/or discussion reported in this paper. 

\subsection{Funding}

Fellowship from ”la Caixa” Foundation (ID 100010434). The fellowship code is LCF/BQ/DI22/11940041. 

Partial financial support from the Agencia Estatal de Investigación and Fondo Europeo de Desarrollo Regional (FEDER, UE) project APASOS (PID2021-122256NB-C21, PID2021-122256NB-C22)

Partial financial support from  María de Maeztu program for Units of Excellence, CEX2021-001164-M funded by  MCIN/AEI/10.13039/501100011033. 

Financial support of the Italian PRIN research project P2022JAYMH "Higher-order complex systems modeling for personalized medicine" funded by NextGenerationEU. 

\subsection{Authors' contributions} 
P.C., A.K. , J.P., and R.N.M. conceived the study, P.C. and R.N.M. wrote scripts analysing data, P.C. and R.N.M.  analysed data, P.C., T.G., A.K., S.M., J.P. and R.N.M.  analysed the results. P.C., T.G., S.M., J.P. and R.N.M. wrote the manuscript. All authors reviewed the manuscript.

\subsection{Acknowledgments}
The project that gave rise to these results received the support of a fellowship from ”la Caixa” Foundation (ID 100010434). The fellowship code is LCF/BQ/DI22/11940041. PC and TG are also grateful for partial financial support from the Agencia Estatal de Investigación and Fondo Europeo de Desarrollo Regional (FEDER, UE) under project APASOS (PID2021-122256NB-C21, PID2021-122256NB-C22), and the María de Maeztu program for Units of Excellence, CEX2021-001164-M funded by  MCIN/AEI/10.13039/501100011033. RNM, SM and JP acknowledge financial support of the Italian PRIN research project P2022JAYMH "Higher-order complex systems modeling for personalized medicine" funded by NextGenerationEU. 

\bibliography{mantegna_epjds}

\newpage

\appendix

\section{Over-expression of disease categories in diseases' communities}
\label{appendix_oe25}

\begin{figure*}
  \centering
    \includegraphics[width=0.8\linewidth]{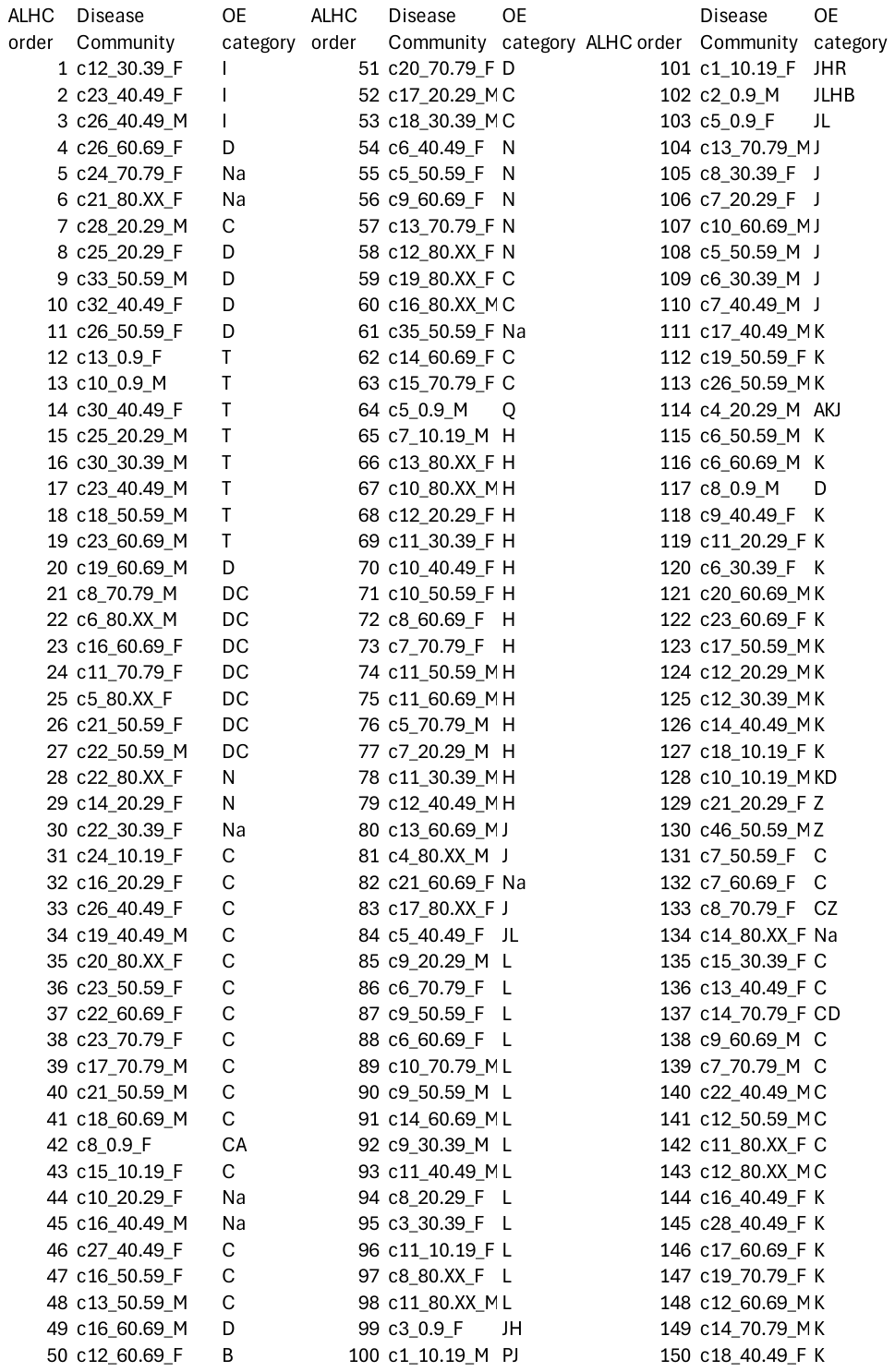}
  \caption{Over-expression of disease categories in diseases communities with more than 25 ICD codes. For each group of three columns, the first is the order of the disease community in the Average Linkage hierarchical tree, the second is the label of the disease community and the third is the over-expressed category (or categories in case of multiple letters). In. the absence of over-expression Na is reported. Average linkage hierarchical clustering (ALHC) order from 1 to 150.}
  \label{fig:OE_A}
\end{figure*}

\begin{figure*}
  \centering
    \includegraphics[width=0.8\linewidth]{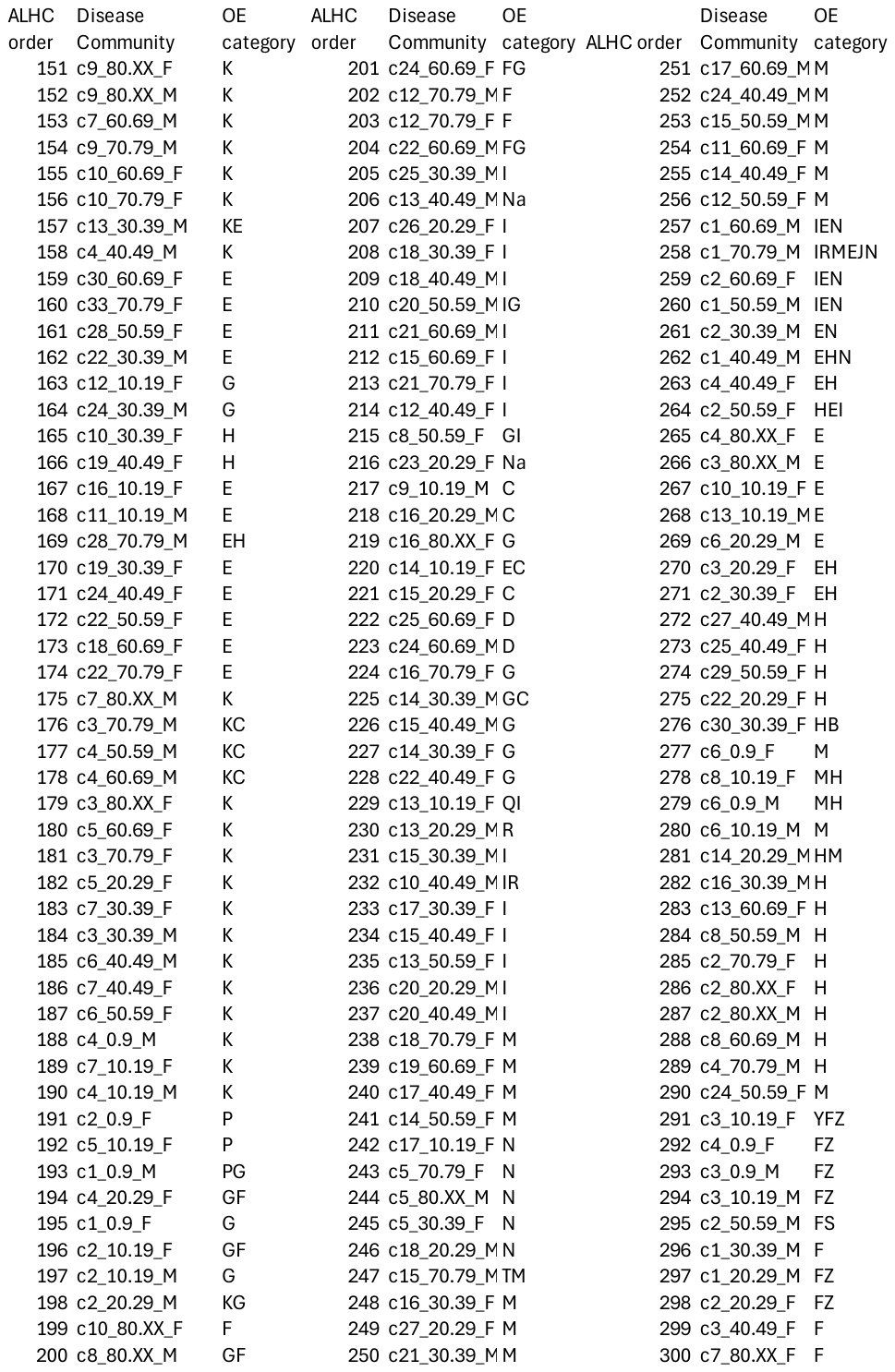}
  \caption{Over-expression of disease categories in disease communities with more than 25 ICD codes. For each group of three columns, the first is the order of the disease community in the Average Linkage hierarchical tree, the second is the label of the disease community and the third is the over-expressed category (or categories in case of multiple letters). In. the absence of over-expression Na is reported. ALHC order from 151 to 300.}
  \label{fig:OE_B}
\end{figure*}

\begin{figure*}
  \centering
    \includegraphics[width=0.8\linewidth]{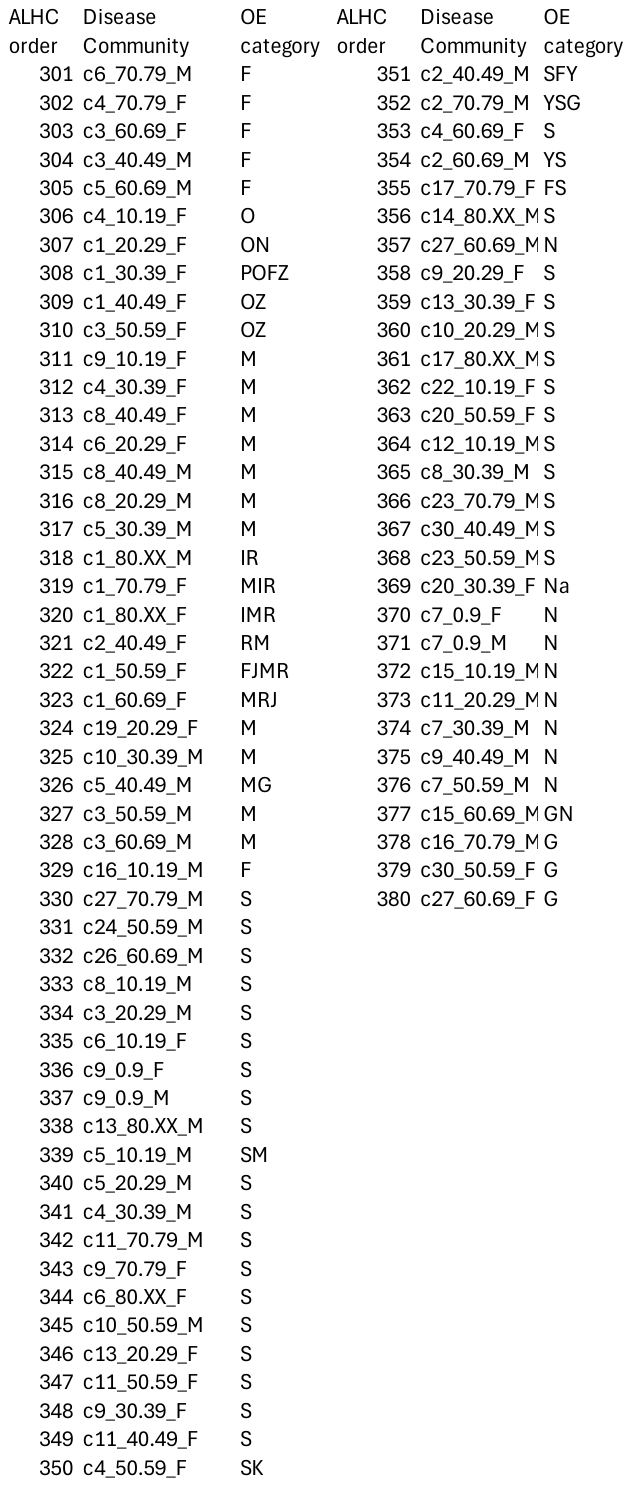}
  \caption{Over-expression of disease categories in diseases communities with more than 25 ICD codes. For each group of three columns, the first is the order of the disease community in the Average Linkage hierarchical tree, the second is the label of the disease community and the third is the over-expressed category (or categories in case of multiple letters). In. the absence of over-expression Na is reported. ALHC order from 301 to 380.}
  \label{fig:OE_C}
\end{figure*}

This appendix shows the full list of over-expression of disease categories in diseases' communities with more than 25 ICD codes. The 380 communities are split into three figures (Fig \ref{fig:OE_A}, \ref{fig:OE_B}, and \ref{fig:OE_C}). Each figure contains three main columns, each listing a series of communities. Each column contains: (i) an integer number (the order of the disease community in the Average Linkage hierarchical tree, then, (ii) the label the community, and (iii) the over-expressed categories of ICD codes in the community. In the absence of over-expression Na is reported.

\section{Jaccard similarity matrix for SVN communities}
\label{appendix_jaccmat}

\begin{figure*}
  \centering
    \includegraphics[width=\linewidth]{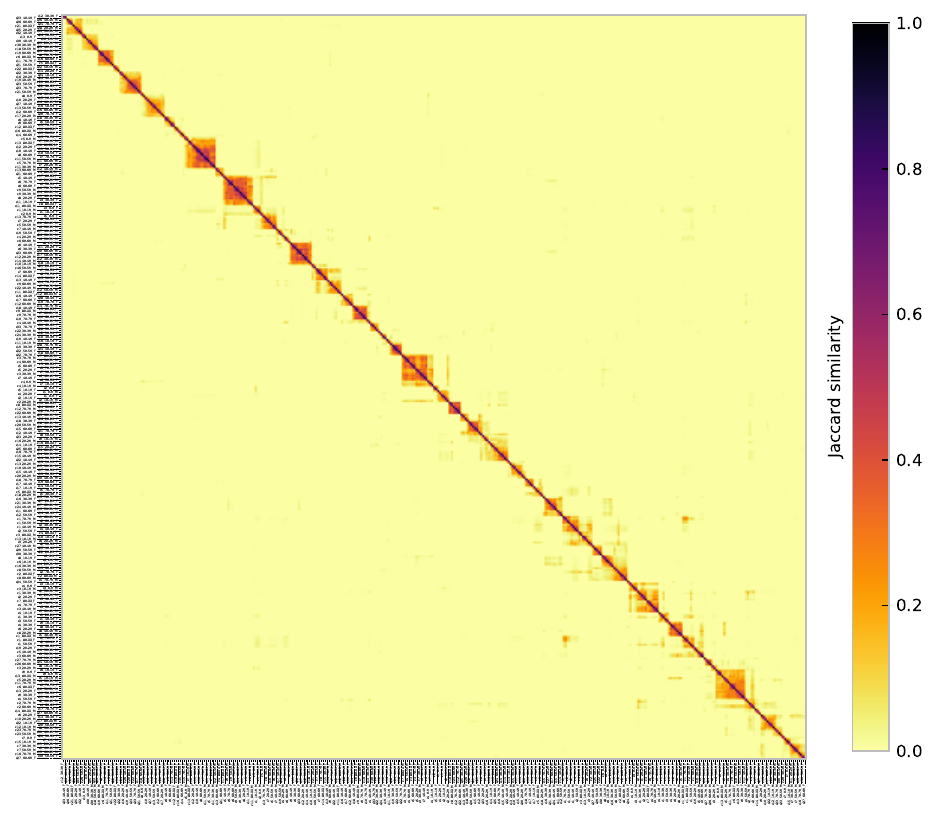}
  \caption{Jaccard similarity matrix among all $380\times 380$ pairs SVN communities. The order of communities in the rows and columns of the matrix is as in the hierarchical tree in Fig.~\ref{fig:HTcolor}.}
\label{fig:HT_Jaccard}
\end{figure*}

This appendix contains the Jaccard similarity matrix among all pairs of 380 SVN communities (Fig. \ref{fig:HT_Jaccard}).

\end{document}